\documentstyle[psfig]{mn}

\newif\ifAMStwofonts

\def\kms{$\rm km\;s^{-1}$}
\def\kmsmpc{$\rm km\;s^{-1}\;Mpc^{-1}$}
\def\ha{H$\alpha$}
\def\h2{H$_{2}$}

\def\hii{H~{\scriptsize II}}
\def\nii{[N~{\scriptsize II}]}
\def\sii{[S~{\scriptsize II}]}
\def\sp{[S~{\scriptsize II}]$\,\lambda6716.5$}
\def\sg{[S~{\scriptsize II}]$\,\lambda6730.9$}
\def\oii{[O~{\scriptsize II}]}
\def\oiii{[O~{\scriptsize III}]$\,\lambda5006.9$}
\def\op{[O~{\scriptsize II}]$\,\lambda3726.2$}
\def\og{[O~{\scriptsize II}]$\,\lambda3728.9$}

\def\cai{Ca~{\scriptsize I}}
\def\caii{Ca~{\scriptsize II}}
\def\lha{$L_{\rm H\alpha}$}

\def\maq{$\rm mag\;arcsec^{-2}$}
\def\ml{$M/L$}
\def\msun{M$_{\odot}$}
\def\lsun{L$_{\odot}$}
\def\lsunV{L$_{\odot\,V}$}
\def\mlV{$M/L_{V}$}
\def\mlsunV{M$_{\odot}$/L$_{\odot\,V}$}

\ifoldfss
  \ifCUPmtlplainloaded \else
    \NewTextAlphabet{textbfit} {cmbxti10} {}
    \NewTextAlphabet{textbfss} {cmssbx10} {}
    \NewMathAlphabet{mathbfit} {cmbxti10} {} 
    \NewMathAlphabet{mathbfss} {cmssbx10} {} 
  \fi
  \ifAMStwofonts
    \ifCUPmtlplainloaded \else
      \NewSymbolFont{upmath} {eurm10}
      \NewSymbolFont{AMSa} {msam10}
      \NewMathSymbol{\upi}     {0}{upmath}{19}
      \NewMathSymbol{\umu}     {0}{upmath}{16}
      \NewMathSymbol{\upartial}{0}{upmath}{40}
      \NewMathSymbol{\leqslant}{3}{AMSa}{36}
      \NewMathSymbol{\geqslant}{3}{AMSa}{3E}

      \let\leq=\leqslant 
      \let\geq=\geqslant 
    \fi
  \fi
\fi 

\ifnfssone
  \newmathalphabet{\mathit}
  \addtoversion{normal}{\mathit}{cmr}{m}{it}
  \addtoversion{bold}{\mathit}{cmr}{bx}{it}
  \newmathalphabet{\mathbfit} 
  \addtoversion{normal}{\mathbfit}{cmr}{bx}{it}
  \addtoversion{bold}{\mathbfit}{cmr}{bx}{it}
  \newmathalphabet{\mathbfss} 
  \addtoversion{normal}{\mathbfss}{cmss}{bx}{n}
  \addtoversion{bold}{\mathbfss}{cmss}{bx}{n}
  \ifAMStwofonts
    \ifCUPmtlplainloaded \else
      \UseAMStwoboldmath
      \makeatletter
      \new@mathgroup\upmath@group
      \define@mathgroup\mv@normal\upmath@group{eur}{m}{n}
      \define@mathgroup\mv@bold\upmath@group{eur}{b}{n}
      \edef\UPM{\hexnumber\upmath@group}
      \new@mathgroup\amsa@group
      \define@mathgroup\mv@normal\amsa@group{msa}{m}{n}
      \define@mathgroup\mv@bold\amsa@group{msa}{m}{n}
      \edef\AMSa{\hexnumber\amsa@group}
      \makeatother
      \mathchardef\upi="0\UPM19
      \mathchardef\umu="0\UPM16
      \mathchardef\upartial="0\UPM40
      \mathchardef\leqslant="3\AMSa36
      \mathchardef\geqslant="3\AMSa3E

      \let\leq=\leqslant 
      \let\geq=\geqslant 
    \fi
  \fi
\fi 

\ifnfsstwo
  \DeclareMathAlphabet{\mathbfit}{OT1}{cmr}{bx}{it}
  \SetMathAlphabet\mathbfit{bold}{OT1}{cmr}{bx}{it}
  \DeclareMathAlphabet{\mathbfss}{OT1}{cmss}{bx}{n}
  \SetMathAlphabet\mathbfss{bold}{OT1}{cmss}{bx}{n}
  \ifAMStwofonts
    \ifCUPmtlplainloaded \else
      \DeclareSymbolFont{UPM}{U}{eur}{m}{n}
      \SetSymbolFont{UPM}{bold}{U}{eur}{b}{n}
      \DeclareSymbolFont{AMSa}{U}{msa}{m}{n}
      \DeclareMathSymbol{\upi}{0}{UPM}{"19}
      \DeclareMathSymbol{\umu}{0}{UPM}{"16}
      \DeclareMathSymbol{\upartial}{0}{UPM}{"40}
      \DeclareMathSymbol{\leqslant}{3}{AMSa}{"36}
      \DeclareMathSymbol{\geqslant}{3}{AMSa}{"3E}

      \let\leq=\leqslant 
      \let\geq=\geqslant 
    \fi
  \fi
\fi 

\ifCUPmtlplainloaded \else
  \ifAMStwofonts \else 
    \def\upi{\pi}
    \def\umu{\mu}
    \def\upartial{\partial}
  \fi
\fi

\title{The kinematics and the origin of the ionized gas in NGC~4036}

\author
[Cinzano et al.]
{P. Cinzano$^1$, H.-W. Rix$^{2,3}$, M. Sarzi$^1$, E.M. Corsini$^4$,
W.W. Zeilinger$^5$ 
\newauthor
and F. Bertola$^1$\\
$^1$ Dipartimento di Astronomia, Universit\`a di Padova,
vicolo dell'Osservatorio 5,  I-35122 Padova, Italy\\
$^2$ Max Planck Institut f{\"u}r Astrophysik,
Karl Schwarzschild Stra{\ss}e 1,  D-85748 Garching bei M{\"u}nchen,
Germany\\
$^3$ Steward Observatory, University of Arizona, Tucson, AZ-85721, USA \\
$^4$ Osservatorio Astrofisico di Asiago, Dipartimento di Astronomia, 
Universit\`a di Padova, via dell'Osservatorio 8,  I-36012 Asiago, Italy\\
$^5$ Institut f{\"u}r Astronomie, Universit{\"a}t Wien,
T{\"u}rkenschanzstra{\ss}e 17, A-1180 Wien, Austria\\}

\date{Accepted ... .
      Received ...;
      in original form ...}

\pagerange{\pageref{firstpage}--\pageref{lastpage}}
\pubyear{}

\begin{document}

\maketitle

\label{firstpage}

\begin{abstract}
We present the kinematics and photometry of the stars and of the
ionized gas near the centre of the S0 galaxy NGC4036. Dynamical models
based on the Jeans Equation have been constructed from the stellar
data to determine the gravitational potential in which the ionized gas
is expected to orbit.  Inside $10''$, the observed gas rotation curve
falls well short of the predicted circular velocity.  Over a
comparable radial region the observed gas velocity dispersion is far
higher than the one expected from thermal motions or small scale
turbulence, corroborating that the gas cannot be following the
streamlines of nearly closed orbits.  We explore several avenues to
understand the dynamical state of the gas: (1) We treat the gas as a
collisionless ensemble of cloudlets and apply the Jeans Equation to
it; this modeling shows that inside $4''$ the gas velocity dispersion
is just high enough to explain quantitatively the absence of rotation.
(2) Alternatively, we explore whether the gas may arise from the `just
shed' mass-loss envelopes of the bulge stars, in which case their
kinematics should simply mimic that of the stars.  he latter approach
matches the data better than (1), but still fails to explain the low
velocity dispersion {\it and\/} slow rotation velocity of the gas for
$5''<r<10''$.  (3) Finally, we explore, whether drag forces on the
ionized gas may aid in explaining its peculiar kinematics.  While all
these approaches provide a much better description of the data than
cold gas on closed orbits, we do not yet have a definitive model to
describe the observed gas kinematics at all radii. We outline
observational tests to understand the enigmatic nature of the ionized
gas.
\end{abstract}


\section{Introduction}
\label{sec_introduction}

Stars and ionized gas provide independent probes of the mass
distribution in a galaxy. The comparison between their kinematics is
particularly important in dynamically hot systems (i.e. whose
projected velocity dispersion is comparable to rotation). In fact in
elliptical galaxies and bulges the ambiguities about orbital
anisotropies can lead to considerable uncertainties in the mass
modeling (e.g. Binney \& Mamon 1982; Rix et al. 1997).

The mass distributions inferred from stellar and gaseous kinematics
are usually in good agreement for discs (where both tracers can be
considered on nearly circular orbits), but often appear discrepant for
bulges (e.g. Fillmore, Boroson \& Dressler 1986; Kent 1988; Kormendy
\& Westpfahl 1989; Bertola et al. 1995b).  There are several
possibilities to explain these discrepant mass estimates in galactic
bulges:

\begin{enumerate}
\item If bulges have a certain degree of triaxiality, depending on
      the viewing angle the gas on closed orbits can either move
      faster or slower than in the `corresponding' axisymmetric case
      (Bertola, Rubin \& Zeilinger 1989). Similarly, the predictions
      of the triaxial stellar models deviate from those in the
      axisymmetric case: whenever $\sigma_{\rm stars} > \sigma_{\rm
      axisym}$, then $v_{\rm stars} < v_{\rm axisym}$;
\item Most of the previous modeling assumes that the gas is dynamically
      cold and therefore rotates at the local circular speed on the
      galactic equatorial plane. If in bulges the gas velocity
      dispersion $\sigma_{\rm gas}$ is not negligible (e.g. Cinzano \&
      van der Marel 1994 hereafter CvdM94; Rix et al. 1995; Bertola et
      al. 1995b), the gas rotates slower than the local circular
      velocity due to its dynamical pressure support.  CvdM94 showed
      explicitly for the E4/S0a galaxy NGC~2974 that the gas and star
      kinematics agree taking into account for the gas velocity
      dispersion. Furthermore, if $\sigma_{\rm gas}$ is comparable to
      the observed streaming velocity, the spatial gas distribution can
      no longer be modeled as a disc;
\item Forces other than gravity (such as magnetic fields, interactions
      with stellar mass loss envelopes and the hot gas component) 
      might act on the ionized gas (e.g. Mathews 1990).
\end{enumerate}

In this paper we pursue the second of these explanations by building
for NGC~4036 dynamical models which take into account both for the
random motions and the three-dimensional spatial distribution of the
ionized gas.

NGC~4036 has been classified S0$_{3}$(8)/Sa in RSA (Sandage
\& Tammann 1981) and S0$^{-}$ in RC3 (de Vaucouleurs et al. 
1991). It is a member of the LGG~266 group, together with NGC~4041,
IC~758, UGC~7009 and UGC~7019 (Garcia 1993). It forms a wide pair with
NGC~4041 with a separation of $17'$ corresponding to 143 kpc at their
mean redshift distance of 29 Mpc (Sandage \& Bedke 1994).  In The
Carnegie Atlas of Galaxies (hereafter CAG) Sandage \& Bedke (1994)
describe it as characterized by an irregular pattern of dust lanes
threaded through the disc in an `embryonic' spiral pattern indicating
a mixed S0/Sa form (see Panel~60 in CAG).  Its total $V-$band apparent
magnitude is $V_T = 10.66$ mag (RC3).  This corresponds to a total
luminosity $L_V = 4.2 \cdot 10^{10}$ \lsunV\ at the assumed distance
of $d = 30.2$ Mpc.  The distance of NGC~4036 was derived as
$d=V_0/H_0$ from the systemic velocity corrected for the motion of the
Sun with respect to the centroid of the Local Group $V_0 = 1509\pm50$
\kms\ (RSA) and assuming $H_0 = 50$ \kmsmpc .  At this distance the
scale is 146 pc arcsec$^{-1}$.

The total masses of neutral hydrogen and dust in NGC~4036 are $M_{\rm
HI} = 1.7 \cdot 10^{9}$ \msun\ and $M_{\rm dust} = 4.4 \cdot 10^{5}$
\msun\ (Roberts et al. 1991). NGC~4036 is known to have emission
lines from ionized gas (Bettoni \& Buson 1987) and the mass of the
ionized gas is $M_{\rm HII} = 7 \cdot 10^{4}$ \msun\ (see
Sec.~\ref{sec_gasmodeling} for a discussion).

This paper is organized as follows.  In Sec.~\ref{sec_observation} we
present the photometrical and spectroscopical observations of
NGC~4036, the reduction of the data, and the analysis procedures to
measure the surface photometry and the major-axis kinematics of stars
and ionized gas. In Sec.~\ref{sec_stars} we describe the stellar
dynamical model (based on Jeans Equations), and we find the potential
due to the stellar bulge and disc components starting from the
observed surface brightness of the galaxy.  In Sec.~\ref{sec_gas} we
use the derived potential to study the dynamics of both the gaseous
spheroid and disc components, assumed to be composed of collisionless
cloudlets orbiting as test particles. In Sec.~\ref{sec_conclusions} we
discuss our conclusions.

\section{Observations and data analysis}
\label{sec_observation}

\subsection{Photometrical observations}
\label{sec_imaging}

\subsubsection{Ground-based data}
\label{sec_ground}

We obtained an image of NGC~4036 of 300 s in the Johnson $V-$band at
the 2.3-m Bok Telescope at Kitt Peak National Observatory on December
22, 1995.

A front illuminated 2048$\times$2048 LICK2 Loral CCD with
$15\times15~\mu$m$^{2}$ pixels was used as detector at the
Richtey-Chretien focus, $f/9$. It yielded a flat field of view with a
$10\farcm1$ diameter. The image scale was $0\farcs43$ pixel$^{-1}$
after a $3\times3$ pixel binning. The gain and the readout noise were
1.8 e$^-$ ADU$^{-1}$ and 8 e$^-$ respectively.

The data reduction was carried out using standard {\sc
IRAF}\footnote{{\sc IRAF} is distributed by the National Optical
Astronomy Observatories which are operated by the Association of
Universities for Research in Astronomy (AURA) under cooperative
agreement with the National Science Foundation} routines. The image
was bias subtracted and then flat-field corrected. The cosmic rays
were identified and removed. A Gaussian fit to field stars in the
resulting image yielded a measurement of the seeing point spread
function (PSF) FWHM of $1\farcs7$.

The sky subtraction and elliptical fitting of the galaxy
i\-so\-pho\-tes were performed by means of the Astronomical Images
Analysis Package ({\sc AIAP}) developed at the Osservatorio
Astronomico di Padova (Fasano 1990).  The sky level was determined by
a polynomial fit to the surface brightness of the frame regions not
contaminated by the galaxy light, and then subtracted out from the
total signal. The isophote fitting was performed masking the bad
columns of the frame and the bright stars of the field. Particular
care was taken in masking the dust-affected regions along the major
axis between $5''$ and $20''$.  No photometric standard stars were
observed during the night. For this reason the absolute calibration
was made scaling the total apparent $V-$band magnitude to $V_{\rm T} =
10.66$ mag (RC3).

\begin{figure}
\centerline{\psfig{figure=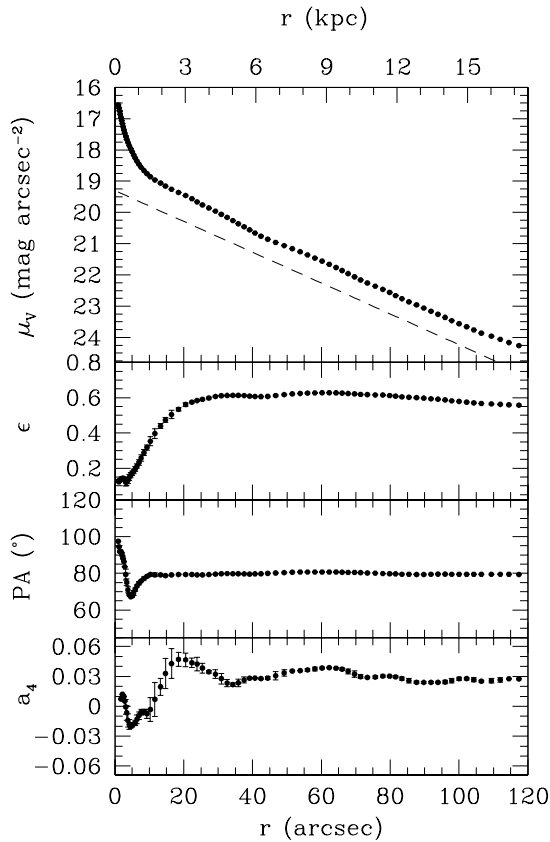,width=12cm,rwidth=9cm,rheight=11.5cm}}
\caption{Ground-based $V-$band observed surface-brightness, ellipticity, 
  position angle and $\cos 4\theta$ coefficient radial profiles of
  NGC~4036. The dashed line in the top panel represents the
  surface-brightness profile of the exponential disc of the best-fit
  kinematical model ($\mu_0 = 19.3$ \maq , $r_d = 22\farcs0$ and $i =
  72\degr$) derived in Sec.~\ref{sec_starmodelingresults}.}    
\label{fig_photometry}
\end{figure}

Fig.~\ref{fig_photometry} shows the $V-$band surface brightness
($\mu_V$), ellipticity ($\epsilon$), major axis position angle (PA),
and the $\cos 4\theta$ ($a_4$) Fourier coefficient of the isophote's
deviations from elliptical as functions of radius along the major
axis.

For $r \la 4''$ the ellipticity is $\ga 0.12$. Between $r\sim4''$ and
$r\sim30''$ it increases to $0.61$. It rises to $0.63$ at $r\sim62''$
and then it decreases to $0.56$ at the farthest observed radius.  The
position angle ranges from $\sim98\degr$ to $\sim67\degr$ in the inner
$5''$. Between $r\sim5''$ and $r\sim10''$ it increases to
$\sim79\degr$ and then it remains constant.  The $\cos 4\theta$
coefficient ranges between $\sim +0.01$ and $\sim -0.02$ for
$r\la5''$. Further out it peaks to $\sim +0.05$ at $r\sim18''$, and
then it decreases to $\sim +0.03$ for $r>30''$.  The abrupt variation
in position angle ($\rm \Delta\,PA\,\simeq\,30\degr$) observed inside
$5''$ leads to an isophote twist that can be interpreted as due to a
slight triaxiality of the inner regions of the stellar bulge.  Anyway
this variation has to be considered carefully due to the presence of
dust pattern in these regions which are revealed by HST imaging (see
Fig.~\ref{fig_hstphotometry}).
 
These results are consistent with previous photometric studies of Kent
(1984) and Michard (1993) obtained in the $r-$ and $B-$band
respectively.  Our measurements of ellipticity follows closely those
by Michard (1993).  Kent (1984) measured the ellipticity and position
angle of NGC~4036 isophotes in $r-$band for $9''\leq\,r\,\leq\,114''$.
Out to $r=78''$ the $r-$band ellipticity is lower than our $V-$band
one by $\sim0.04$. The $r-$band position angle profile differs from
our only at $r\sim16''$ ($\epsilon_r=76\degr$) and for $r>78''$
($\epsilon_r=81\degr$). We found $\cos 4\theta$ isophote deviation
from ellipses to have a radial profile in agreement with that by
Michard (1993).

\subsubsection{Hubble Space Telescope data}
\label{sec_hst}

In addition, we derived the ionized gas distribution in the nuclear
regions of NGC~4036 by the analysis of two Wide Field Planetary Camera
2 (WFPC2) images which were extracted from the {\it Hubble Space
Telescope\/} archive\footnote{Observations with the NASA/ESA {\it
Hubble Space Telescope\/} were obtained from the data archive at the
Space Telescope Science Institute (STScI), operated by AURA under NASA
contract NAS 5-26555.}.

We used a 300 s image obtained on August 08, 1994 with the F547M
filter (principal investigator: Sargent GO-05419) and a 700 s image
taken on May 15, 1997 with the F658N filter (principal investigator:
Malkan GO-06785).

\begin{figure}
\vspace{-2.5cm}
\centerline{\psfig{figure=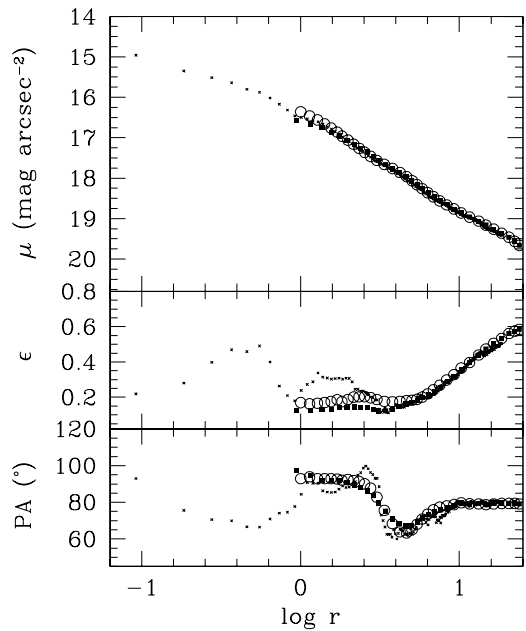,width=12cm,rwidth=9cm,rheight=11cm}}
\caption{Surface brigthness, ellipticity and position angle radial   
  profiles of the nuclear regions of NGC~4036 obtained from a HST
  image (crosses) compared to the ground-based data before (filled
  squares) and after (open circles) the seeing deconvolution (see
  Sec.~\ref{sec_deconvolution}).  The WFPC2 image is in the
  F547W$-$band while the ground-based ones are in the $V-$band.}
\label{fig_hstphotometry}
\end{figure}

The standard reduction and calibration of the images were performed at
the STScI using the pipeline-WFPC2 specific calibration algorithms.
Further processing using the IRAF STSDAS package involved the cosmic
rays removal and the alignment of the images (which were taken with
different position angles).  The surface photometry of the F547M image
was carried out using the STSDAS task ELLIPSE without masking the dust
lanes.  In Fig.~\ref{fig_hstphotometry} we plot the resulting
ellipticity ($\epsilon$) and major axis position angle (PA) of the
isophotes as functions of radius along the major axis.

\begin{figure}
\vspace{-2.5cm}
\centerline{\psfig{figure=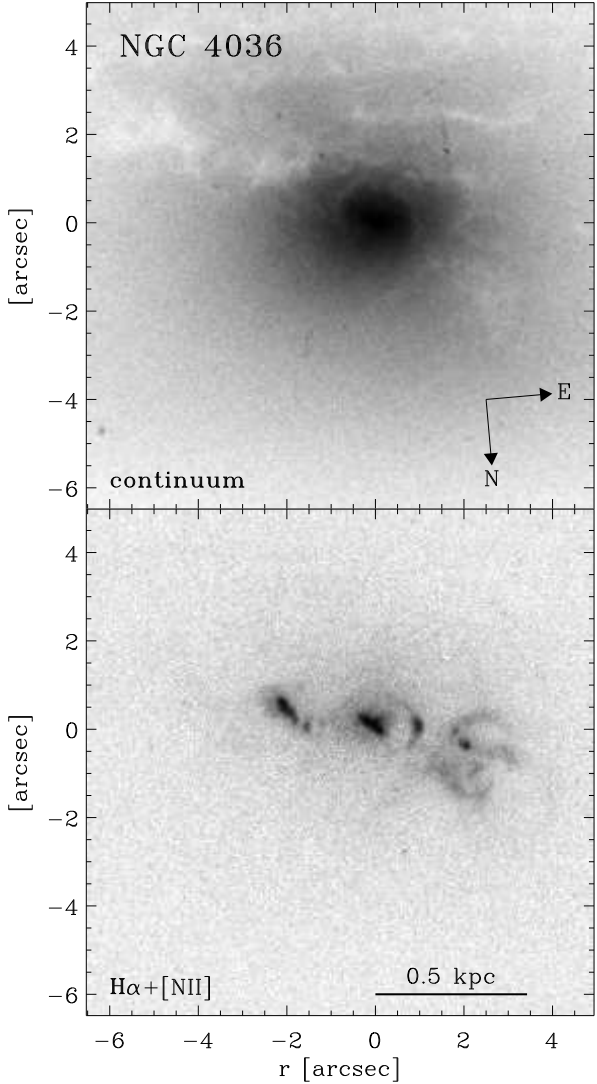,width=12cm,rwidth=9cm,rheight=17cm}}
\caption{WFPC2 stellar continuum image (upper panel) and the 
  con\-ti\-nuum-subtracted \ha$+$\nii\ emission image (lower panel) of the
  nucleus of NGC~4036.}
\label{fig_halpha}
\end{figure}

The continuum-free image of NGC~4036 (Fig.~\ref{fig_halpha}) was
obtained by subtracting the continuum-band F547M image suitably
scaled, from the emission-band F658N image The mean scale factor for
the continuum image was estimated by comparing the intensity of a
number of 5$\times$5 pixels regions near the edges of the frames in
the two bandpasses. These regions were chosen in the F658N image to be
emission free. Our continuum-free image reveals that less than the
$40\%$ of the \ha$\times$\nii\ flux of NGC~4036 derives from a clumpy
structure of about $6'' \times 2''$.  The center of this complex
filamentary structure which is embedded in a smooth emission pattern
coincides with the position of the maximum intensity of the continuum.

\subsection{Spectroscopical observations}
\label{sec_spectroscopy}

A major-axis (PA$=85\degr$) spectrum of NGC~4036 was obtained on March
30, 1989 with the Red Channel Spectrograph at the Multiple Mirror
Telescope\footnote{The MMT is a joint facility of the Smithsonian
Institution and the University of Arizona.} as a part of a larger
sample of 8 S0 galaxies (Bertola et al. 1995b).

The exposure time was 3600 s and the 1200 grooves mm$^{-1}$ grating
was used in combination with a $1\farcs25\times 180''$ slit. It
yielded a wavelength coverage of 550 \AA\ between about 3650 \AA\ and
about 4300
\AA\ with a reciprocal dispersion of 54.67 \AA~mm$^{-1}$.  The spectral
range includes stellar absorption features, such as the the \caii\ H
and K lines ($\lambda\lambda\,$3933.7, 3968.5 \AA) and the \cai\
g-band ($\lambda\,$4226.7 \AA), and the ionized gas \oii\ emission
doublet ($\lambda\lambda\,$3726.2. 3728.9 \AA). The instrumental
resolution was derived measuring the $\sigma$ of a sample of single
emission lines distributed all over the spectral range of a comparison
spectrum after calibration. We checked that the measured $\sigma$'s
did not depend on wavelength, and we found a mean value $\sigma = 1.1$
\AA .  It corresponds to a velocity resolution of $\sim 88$ \kms\ at
3727 \AA\ and $\sim 83$
\kms\ at 3975 \AA .  The adopted detector was the 800$\times$800 Texas
Instruments CCD, which has 15$\times$15\ $\mu$m$^{2}$ pixel size.  No
binning or rebinning was done. Therefore each pixel of the frame
corresponds to $0.82$~\AA~$\times\ 0\farcs33$.

Some spectra of late-G and early-K giant stars were taken with the
same instrumental setup for use as velocity and velocity dispersion
templates in measuring the stellar kinematics.  Comparison
helium-argon lamp exposures were taken before and after every object
integration.  The seeing FWHM during the observing night was in
between $1''$ and $1\farcs5$.

The data reduction was carried out with standard procedures from the
{\sc ESO-MIDAS}\footnote{{\sc MIDAS} is developed and mantained by the
European Southern Ob\-servatory} package. The spectra were bias
subtracted, flat-field corrected, cleaned for cosmic rays and
wavelength calibrated. The sky contribution in the spectra was
determined from the edges of the frames and then subtracted.

\subsubsection{Stellar kinematics}
\label{sec_starkinematics}

The stellar kinematics was analyzed with the Fourier Quotient Method
(Sargent et al. 1977) as applied by Bertola et al. (1984).  The K4III
star HR~5201 was taken as template. It has a radial velocity of $-2.7$
\kms\ (Evans 1967) and a rotational velocity of 10 \kms\ (Bernacca \&
Perinotto 1970). No attempt was made to produce a master template by
combining the spectra of different spectral types, as done by Rix \&
White (1992) and van der Marel et al. (1994). The template spectrum
was averaged along the spatial direction to increase the
signal-to-noise ratio ($S/N$).  The galaxy spectrum was rebinned along
the spatial direction until a ratio $S/N \geq 10$ was achieved at each
radius.  Then spectra of galaxy and template star were rebinned to a
logarithmic wavelength scale, continuum subtracted and endmasked.  The
least-square fitting of Gaussian broadened spectrum of the template
star to the galaxy spectrum was done in the Fourier space over the
restricted range of wavenumbers $\left[k_{\it min},k_{\it
max}\right]=\left[5,200\right]$. In this way we rejected the
low-frequency trends (corresponding to $k<5$) due to the residuals of
continuum subtraction and the high-frequency noise (corresponding to
$k>200$) due to the instrumental resolution. (The wavenumber range is
important in particular in the Fourier fitting of lines with
non-Gaussian profiles, see van der Marel \& Franx 1993; CvdM94).
 
The values obtained for the stellar radial velocity and velocity
dispersion as a function of radius are given in
Tab.~\ref{tab_stars}. The table reports the galactocentric distance
$r$ in arcsec (Col.~1), the heliocentric velocity $V$ (Col.~2) and its
error $\delta V$ (Col.~3) in \kms , the velocity dispersion $\sigma$
(Col.~4) and its error $\delta\sigma$ (Col.~5) in \kms .  The values
for the stellar $\delta V$ and $\delta\sigma$ are the formal errors
from the fit in the Fourier space.

\begin{figure}
\centerline{\psfig{figure=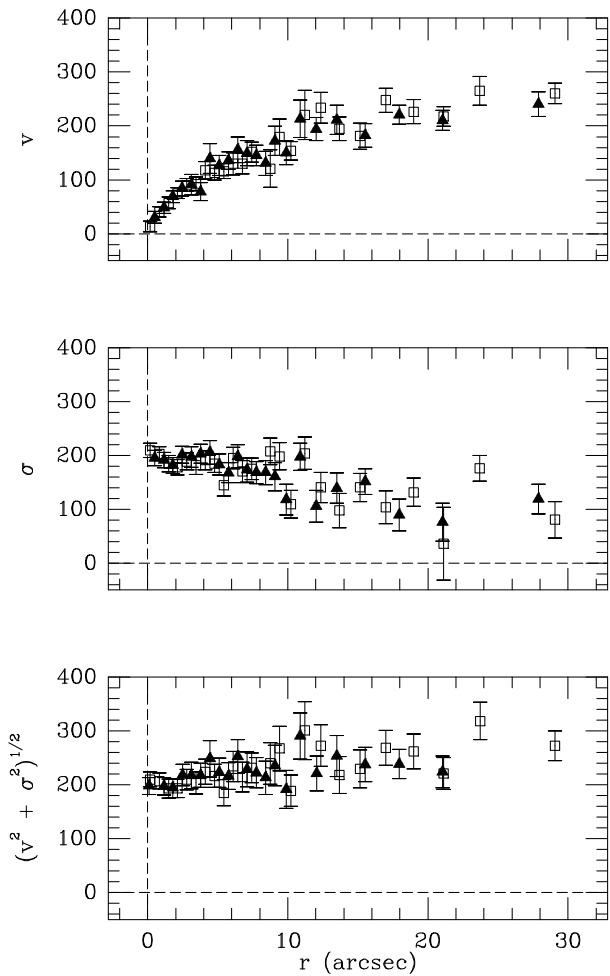,width=8cm}}
\caption{Stellar kinematics along the major axis of NGC~4036.
 The folded rotation velocity curve (top panel), velocity dispersion
 profile (middle panel) and rms velocity curve (bottom panel)
 are shown in \kms . Open squares and filled triangles represent data
 derived for the approaching W and receding E side respectively.}
\label{fig_stars}
\end{figure}

The systemic velocity was subtracted from the observed heliocentric
velocities and the profiles were folded about the centre, before
plotting.  We derive for the systemic heliocentric velocity a value
$V_\odot = 1420 \pm 15$ \kms . Our determination is in agreement
within the errors with $V_\odot = 1397 \pm 27$ \kms\ (RC3) and
$V_\odot = 1382 \pm 50$ \kms\ (RSA) derived from optical observations
too.  The resulting rotation curve, velocity dispersion profile and
rms velocity ($\sqrt{v^2+\sigma^2}$) curve for the stellar component
of NGC~4036 are shown in Fig.~\ref{fig_stars}. The kinematical
profiles are symmetric within the error bars with respect to the
galaxy centre.  For $r \la 2''$ the rotation velocity increases almost
linearly with radius up to $\sim 100$ \kms , remaining approximatively
constant between $2''$ and $4''$. Outwards it rises to the farthest
observed radius.  It is $\sim 180$ \kms\ at $9''$, $\sim 220$ \kms\ at
$21''$ and $\sim 260$ \kms\ at $29''$.  The velocity dispersion
$\sigma \sim 210$ \kms\ in the centre and at $r\sim4''$ with a `local
minimum of $\sim 180$ at $r=2''$. Further out it declines to values
$\la 120$ \kms .

\begin{figure}
\centerline{\psfig{figure=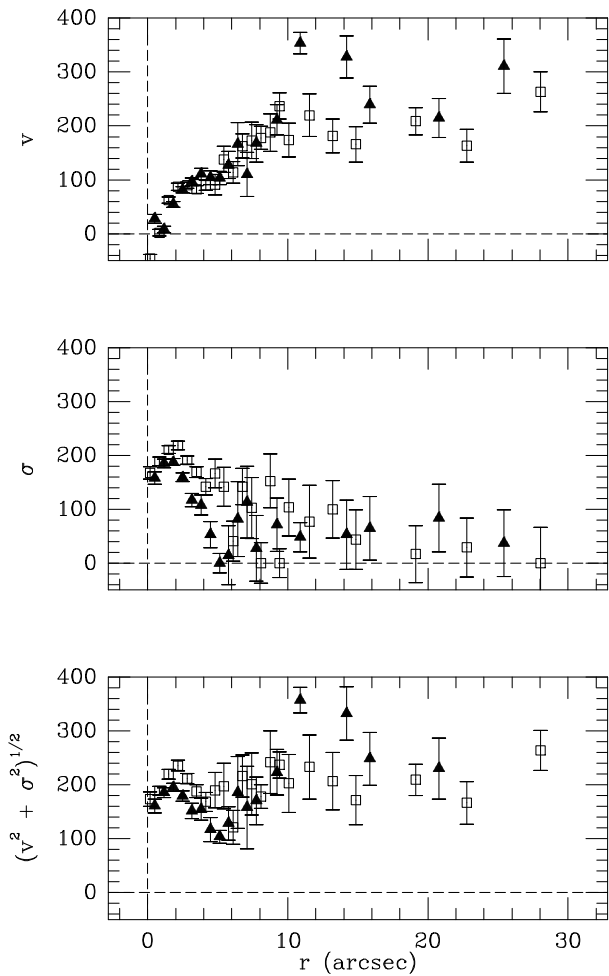,width=8cm}}
\caption{Ionized gas kinematics along the major axis of NGC~4036.
 The folded rotation velocity curve (top panel), velocity dispersion
 profile (middle panel) and rms velocity curve (bottom panel) are
 shown in \kms . Open squares and filled triangles represent data
 derived for the approaching W and receding E side respectively.  The
 velocity error bars indicate the formal error of double Gaussian fit
 to the \oii\ doublet.  The error bars in the velocity dispersions
 take also into account for the subtraction of the instrumental
 dispersion.} 
\label{fig_gas}
\end{figure}

\subsubsection{Ionized gas kinematics}
\label{sec_gaskinematics}

To determine the ionized gas kinematics we studied the
\oii\ ($\lambda\lambda\,$3726.2, 3728.9 \AA) emission doublet.
In our spectrum the two lines are not resolved at any radius.  We
obtained smooth fits to the \oii\ doublet using a two-steps procedure.
In the first step the emission doublet was analyzed by fitting a
double Gaussian to its line profile, fixing the ratio between the
wavelengths of the two lines and assuming that both lines have the
same dispersion.  The intensity ratio of the two lines depends on the
state (i.e.  electron density and temperature) of the gas
(e.g. Osterbrock 1989). We found a mean value of \op/\og$\, =
0.8\pm0.1$ without any significative dependence on radius.  The
electron density derived from the obtained intensity ratio of
\oii\ lines is in agreement with that derived (at any assumed electron
temperature, see Osterbrock 1989) from the intensity ratio of the
\sii\ lines (\sp/\sg=1.23) found by Ho, Filippenko \& Sargent (1997).
In the second step we fitted the line profile of the emission doublet
by fixing the intensity ratio of its two lines at the value above.  At
each radius we derived the position, the dispersion and the
uncalibrated intensity of each \oii\ emission line and their formal
errors from the best-fitting double Gaussian to the doublet plus a
polynomial to its surrounding continuum.  The wavelength of the lines'
centre was converted into the radial velocity and then the
heliocentric correction was applied. The lines' dispersion was
corrected for the instrumental dispersion and then converted into the
velocity dispersion.

The measured kinematics for the gaseous component in NGC~4036 is given
in Tab.~\ref{tab_gas}. The table contains the galactocentric distance
$r$ in arcsec (Col.~1), the heliocentric velocity $V$ (Col.~2) and its
error $\delta V$ (Col.~3) in \kms\ the velocity dispersion $\sigma$
(Col.~4) and its errors $\delta\sigma_+$ and $\delta\sigma_-$ (Cols.~5
and 6) in \kms .  The gas velocity errors $\delta V$ are the formal
errors for the double Gaussian fit to the \oii\ doublet. The gas
velocity dispersion errors $\delta\sigma_+$ and $\delta\sigma_-$ take
also account for the subtraction of the instrumental dispersion.

\begin{figure}
\vspace{-1.5cm}
\centerline{\psfig{figure=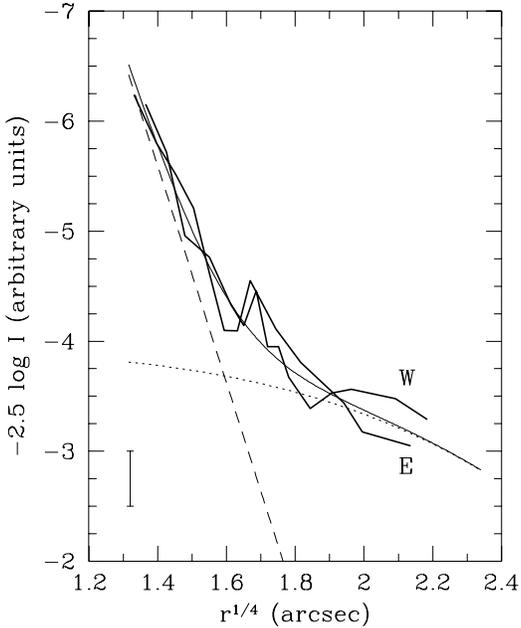,width=12cm,rwidth=9cm,rheight=10cm}}
\caption{The intensity of the \op\ line as function of radius in NGC 4036.
 Since the spectrum was not flux calibrated the scale has an arbitrary
 zero point.  The thick continuous lines connect the measurements
 along the W and E side respectively.  The vertical bar on the
 left-hand side of the panel represents the typical error for the
 data.  The dashed and dotted lines represent respectively the profile
 of the gaseous $R^{1/4}$ spheroid and gaseous exponential disc
 derived in Sec.~\ref{sec_gasmodeling}. The thin continuous line is
 the fit to the data.} 
\label{fig_intensity}
\end{figure}

The rotation curve, velocity dispersion profile and rms velocity curve
for the ionized gas component of NGC~4036 resulting after folding
about the centre are shown Fig.~\ref{fig_gas}.  The \op\ intensity
profile as a function of radius is plotted in
Fig.~\ref{fig_intensity}.  The gas rotation tracks the stellar
rotation remarkably well. They are consistent within the errors to one
another.  The gas velocity dispersion has central dip of $\sim 160$
\kms\ with a maximum of $\sim 220$ \kms\ at $r\sim2''$. It remains
higher than 100 \kms\ up to $\sim4''$ before decreasing to lower
values. The velocity dispersion profile appears to be less symmetric
than the rotation curve.  Indeed between $2''$ and $5''$ the velocity
dispersion measured onto the E side rapidly drops from its observed
maximum to $\sim 50$ \kms, while in the W side it smoothly declines to
$\sim 140$ \kms. Errors on the gas velocity dispersion increase at
large radii as the gas velocity dispersion becomes comparable to the
instrumental dispersion.

\subsubsection{Comparison with kinematical data by Fisher (1997)}
\label{sec_comparisonfisher}

The major-axis kinematics for stars and gas we derived for NGC~4036
are consistent within the errors with the measurements of Fisher
(1997, hereafter F97). The only exception is represented by the
differences of $20\%$--$30\%$ between our and F97 stellar velocity
dispersions in the central $8''$. In these regions F97 finds a flat
velocity dispersion profile with a plateau at $\sigma \sim 170$ \kms
. To measure stellar kinematics he adopted the Fourier Fitting Method
(van der Marel \& Franx 1993) directly on the line-of-sight velocity
distribution derived with Unresolved Gaussian Decomposition Method
(Kuijken \& Merrifield 1993).  For $|r| \la 8''$ the NGC~4036 line
profiles are asymmetric (displaying a tail opposite to the direction
of rotation) and flat-toped, as result from the $h_3$ and $h_4$ radial
profiles.  The $h_3$ and $h_4$ parameters measure respectively the
asymmetric and symmetric deviations of the line profile from a
Gaussian (van der Marel \& Franx 1993; Gerhard 1993).  For NGC~4036
the $h_3$ term anticorrelates with $v$, rising to $\sim +0.1$ in the
approaching side and falling to $\sim -0.1$ in the receding side. The
$h_4$ term exhibits a negative value ($\sim-0.03$).

\section{Modeling the stellar kinematics}
\label{sec_stars}

\subsection{Modeling technique}
\label{sec_starmodeling}

We built an axisymmetric bulge-disc dynamical model for NGC~4036
applying the Jeans modeling technique introduced by Binney, Davies \&
Illingworth (1990), developed by van der Marel, Binney \& Davies
(1990) and van der Marel (1991), and extended to two-component
galaxies by CvdM94 and to galaxies with a DM halo by Cinzano (1995)
and Corsini et al. (1998).  For details the reader is referred the
above references.

The main steps of the adopted modeling are (i) the calculation of the
bulge and disc contribution to the potential from the observed surface
brightness of NGC~4036; (ii) the solution of the Jeans Equations to
obtain separately the bulge and disc dynamics in the total potential;
and (iii) the projection of the derived dynamical quantities onto the
sky-plane taking into account seeing effects, instrumental set-up and
reduction technique to compare the model predictions with the measured
stellar kinematics. In the following each invidual step is briefly
discussed:

\begin{enumerate}

\item We model NGC 4036 with an infinitesimally thin exponential disc in
      its equatorial plane. The disc surface mass density is specified
      for any inclination $i$, central surface brightness $\mu_0$,
      scale length $r_d$ and constant mass-to-light ratio $(M/L)_d$.
      The disc potential is calculated from the surface mass density
      as in Binney \& Tremaine (1987).  The limited extension of our
      kinematical data (measured out $r<30''\,\simeq\,0.5\,R_{\it
      opt}$\footnote{The optical radius $R_{\it opt}$ is the radius
      encompassing the $83\%$ of the total integrated light.})
      prevents us to disentangle in the assumed constant mass-to-light
      ratios the possible contribution of a dark matter halo.\\
      The surface brightness of the bulge is obtained by subtracting
      the disc contribution from the total observed surface
      brightness.  The three-dimensional luminosity density of the
      bulge is obtained deprojecting its surface brightness with an
      iterative method based on the Richardson-Lucy algorithm
      (Richardson 1972; Lucy 1974).  Its three-dimensional mass
      density is derived by assuming a constant mass-to-light ratio
      $(M/L)_b$.  The potential of the bulge is derived solving the
      Poisson Equation by a multipole expansion (e.g. Binney \&
      Tremaine 1987).

\item The bulge and disc dynamics are derived by separately solving the
      Jeans Equations for each component in the total potential of the
      galaxy. For both components we assume a two integral
      distribution function of the form $f=f(E,L_z)$.  It implies that
      the vertical velocity dispersion $\sigma^2_z$ is equal to second
      radial velocity moment $\sigma_R^2$ and that
      $\overline{v_Rv_z}=0$.  Therefore the Jeans Equations becomes a
      closed set, which can be solved for the unknowns
      $\overline{v^2_{\phi}}$ and $\sigma^2_R = \sigma^2_z$.  Other
      assumptions to close the Jeans Equations are also possible (e.g.
      van der Marel \& Cinzano 1992).\\
      For the bulge we made the same hypotheses of Binney et
      al. (1990). A portion of the second velocity moment
      $\overline{v^2_{\phi}}$ is assigned to bulge streaming velocity
      $\overline{v_\phi}$ following Satoh's (1980) prescription.\\ %
      For the disc we made the same hypotheses of Rix \& White (1992)
      and CvdM94.  The second radial velocity moment $\sigma^2_R$ in
      the disc is assumed to fall off exponentially with a scale
      length $R_{\sigma}$ from a central value of $\sigma^2_{d0}$.
      The azimuthal velocity dispersion $\sigma^2_{\phi}$ in the disc
      is assumed to be related to $\sigma^2_R$ according to the
      relation from epicyclic theory [cfr. Eq.~(3-76) of Binney \&
      Tremaine (1987)].  As pointed out by CvdM94 this relation may
      introduce systematic errors (Kuijken \& Tremaine 1992; Evans \&
      Collett 1993; Cuddeford \& Binney 1994).  The disc streaming
      velocity $\overline{v_{\phi}}$ (i.e. the circular velocity
      corrected for the asymmetric drift) is determined by the Jeans
      equation for radial equilibrium. 

\item We projected back onto the plane of the sky (at the given
      inclination angle) the dynamical quantities of both the bulge
      and the disc to find the line-of-sight projected streaming
      velocity and velocity dispersion.  We assumed that both the
      bulge and disc have a Gaussian line profile. At each radius
      their sum (normalized to the relative surface brightness of the
      two components) represents the model-predicted line profile.  As
      in CvdM94 the predicted line profiles were convolved with the
      seeing PSF of the spectroscopic observations and sampled over
      the slit width and pixel size to mimic the observational
      spectroscopic setup.  We mimicked the Fourier Quotient method
      for measuring the stellar kinematics fitting the predicted line
      profiles with a Gaussian in the Fourier space to derive the
      line-of-sight velocities and velocity dispersions for the
      comparison with the observed kinematics.  The problems of
      comparing the true velocity moments with the Fourier Quotient
      results were discussed by van der Marel \& Franx (1993) and by
      CvdM94.
\end{enumerate}

\subsection{Results for the stellar component}
\label{sec_starresults}

\subsubsection{Seeing-deconvolution}
\label{sec_deconvolution}
 
The modeling technique described in Sec.~\ref{sec_starmodeling}
derives the three-dimensional mass distribution from the
three-dimensional luminosity distribution inferred from the observed
surface photometry by a fine-tuning of the disc parameters leading to
the best fit on the kinematical data.  We performed a
seeing-deconvolution of the $V-$band image of NGC~4036 to take into
account the seeing effects on the measured photometrical quantities
(surface-brightness, ellipticity and $\cos4\theta$ deviation profiles)
used in the deprojection of the two-dimensional luminosity
distribution.  We obtained a restored NGC~4036 image through an
iterative method based on the Richardson-Lucy algorithm (Richardson
1972; Lucy 1974) available in the IRAF package STSDAS. We assumed the
seeing PSF to be a circular Gaussian with a FWHM$=1\farcs7$.  Noise
amplification represents a main backdraw in all Richardson-Lucy
iterative algorithms and the number of iterations needed to get a good
image restoration depends on the steepness of the surface-brightness
profile (e.g. White 1994).  After 6 iterations, we did not notice any
more substantial change in the NGC~4036 surface-brightness profile
while the image became too noisy. Therefore we decided to stop at the
sixth iteration.  The surface-brightness, ellipticity, position angle
radial profiles of NGC~4036 after the seeing-deconvolution are
displayed in Fig.~\ref{fig_hstphotometry} compared with the HST and
the unconvolved ground-based photometry.  The raises found at small
radii for the surface brightness ($r\sim 0.2$ \maq) and for the
ellipticity ($r\sim 0.05$) are comparable in size to the values found
by Peletier et al. (1990) for seeing effects in the centre of
ellipticals.

\subsubsection{Bulge-disc decomposition}
\label{sec_decomposition}

We performed a standard bulge-disc decomposition with a parametric fit
(e.g. Kent 1985) in order to find a starting guess for the exponential
disc parameters to be used in the kinematical fit.  We decomposed the
seeing-deconvolved surface-brightness profile on both the major and
the minor axis as the sum of an $R^{1/4}$ bulge, having
surface-brightness profile
\begin{equation}
\mu_b(r) = \mu_e + 8.3268 \left[\left(\frac{r}{r_e}\right)^{1/4}-1
\right]\,,
\end{equation}
plus an exponential disc, having surface-brightness profile
\begin{equation}
\mu_d(r) = \mu_0 + 1.0857 \left(\frac{r}{r_d}\right)\,.
\end{equation}
We assumed that the minor-axis profile of each component is the same
as the major-axis profile, but scaled by a factor $1-\epsilon = b/a$.
A least-squares fit of the photometric data provides $\mu_e$, $r_e$
and $\epsilon$ of the bulge, $\mu_0$, $r_d$ of the disc, and the
galaxy inclination $i$ (see Tab.~\ref{tab_decomposition} for the
results).

Kent (1984) measured surface photometry of NGC~4036 in the $r-$band.
He decomposed the major- and minor-axis profiles in an $R^{1/4}$ bulge
and in an exponential disc (Kent 1985).  A rough but useful comparison
between the bulge and disc parameters resulting from the two
photometric decompositions is possible by a transformation from $r$-
to $V-$band of Kent's data. We derived from Kent's surface brightness
profile along the major axis of NGC~4036 its (extrapolated) total
magnitude $r_T = 10.56\pm0.02$ corresponding to $V_T-r_T =
0.13\pm0.11$.  Then we converted the $\mu_0$ and $\mu_e$ values from
the $r$- to the $V-$band (see Tab.~\ref{tab_decomposition}).  The
differences between the best-fit parameters obtained from the two
decompositions are lower than $10\%$ except for bulge ellipticity (our
value is $\sim64\%$ than the Kent's one).  These discrepancies are
consistent with the differences in the slope of the two
surface-brightness profiles, with differences between the Kent's
approach to correct for the seeing effects [Kent (1985) convolved the
theoretical bulge and disc profiles with the observed Gaussian seeing
profile] and with the uncertainties of the conversion from $r$- to
$V-$band of Kent's surface brightnesses.

\begin{table}
\caption{The bulge-disc decomposition parameters}
\label{tab_decomposition}
\begin{tabular}{lcccccc}
\multicolumn{1}{c}{} & \multicolumn{3}{c}{bulge} &
\multicolumn{3}{c}{disc}  \\
           &$\mu_e$& $r_e$      & $\epsilon$& $\mu_0$& $r_d$      & $i$
           \\
this paper &20.4   & $12\farcs8$& 0.11      & 18.7   & $22\farcs1$&
$74\fdg9$\\
Kent (1985)&20.7   & $13\farcs1$& 0.07      & 18.6   & $21\farcs0$&
$71\fdg9$\\
\end{tabular}

\medskip
Note. $\mu_e$ and $\mu_0$ are given in $V-$\maq . 
Kent's (1984) surface brightnesses have been converted from the $r$- to 
the $V-$band.
\end{table}

\subsubsection{Modeling results}
\label{sec_starmodelingresults}

\begin{figure}
\centerline{\psfig{figure=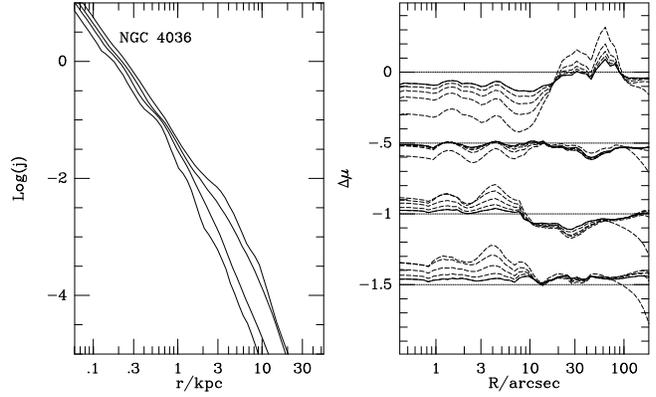,width=8.5cm,angle=270}}
\caption {The deprojection of the surface brightness of the
  spheroidal component of NGC~4036 for the best-fit kinematical model.
  The right panel shows the residual surface brightness $\Delta\mu$ in
  \maq\ after each Lucy iteration (dashed lines) from an initial Jaffe
  fit to the actual NGC~4036 bulge brightness.  The residuals are
  shown for four axes (major through minor axis: top through bottom).
  For each set of curves: the solid line corresponds to the projected
  adopted model for the 3-D luminosity density and the dotted line
  corresponds to a perfect deprojection. At each interaction if the
  model is brighter than the galaxy, the corresponding dashed or
  continuous curve is below the dotted line. In the left panel the
  final three-dimensional luminosity density profile of the spheroidal
  component of NGC~4036 right panel is given in units of $10^{10} {\rm
  L_{\odot}\;kpc^{-3}}$ for the same four axes (minor through major
  axis: innermost through outermost curve).} 
\label{fig_deprojection}
\end{figure}

We looked for the disc parameters leading to the best fit of the
observed stellar kinematics, using as starting guess those resulting
from our best-fit photometric decomposition, with $|\mu_0 -
1|\leq18.7$ \maq , $|r_d - 5''|\leq22\farcs1$, and
$|i-5\degr|\leq74\fdg9$.  For any exponential disc of fixed $\mu_0,
r_d, i$ (in the investigated range of values), we subtracted its
surface brightness from the total seeing-deconvolved surface
brightness. The residuum surface brightness was considered to be
contributed by the bulge. Being $\mu_0$, $r_d$ and $i$ correlated, a
bulge component with a surface-brightness profile consistent with that
resulting from the photometric decomposition, was obtained only by
taking exponential discs characterized by large $\mu_0$ values in
combination with large values of $r_d$ (i.e. larger but fainter discs)
or by small $\mu_0$ and small $r_d$ (i.e. smaller but brighter discs).
After subtracting the disc contribution to the total surface
brightness, the three-dimensional luminosity density of the bulge has
been obtained after seven Richardson-Lucy iterations starting from a
fit to the actual bulge surface brightness with the projection of a
flattened Jaffe profile (1983).  The residual surface brightness
($\Delta\mu = \mu_{\rm model}-\mu_{\rm obs}$) after each iteration and
the final three-dimensional luminosity density profiles of the
spheroidal component of NGC~4036 along the major, the minor and two
intermediate axis is plotted for the kinematical best-fit model in
Fig.~\ref{fig_deprojection} up to $100''$ (the stellar and ionized-gas
kinematics are measured to $\sim30''$).

\begin{figure}
\centerline{\psfig{figure=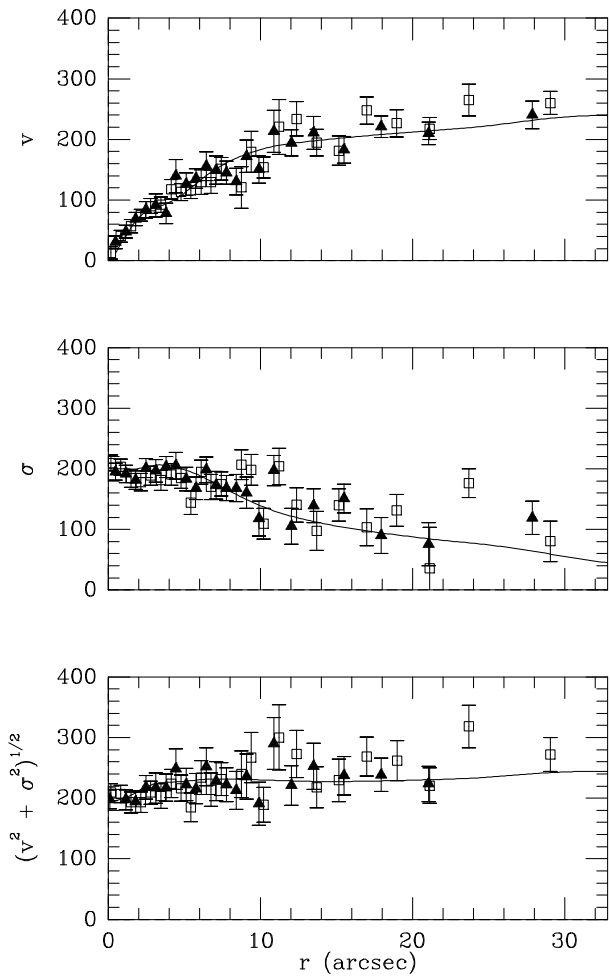,width=8cm}}
\caption{Comparison of the model predictions to the stellar major-axis
  kinematic for NGC 4036. Data points are as in Fig.~\ref{fig_stars}.
  The solid curves represent the velocity (upper panel), velocity
  dispersion (middle panel) and rms velocity (lower panel) radial
  profiles of the best-fit model (in which the disc parameters are
  $\mu_0 = 19.3$ \maq , $r_d = 22\farcs0$ and $i=72\degr$).}
\label{fig_starmodel}
\end{figure}

We applied the modeling technique described in
Sec.~\ref{sec_starmodeling} considering only models in which the bulge
is an oblate isotropic rotator (i.e. $k=1$) and in which the bulge and
the disc have the same mass-to-light ratio $(M/L)_b=(M/L)_d$. The \ml\
determines the velocity normalization and was chosen (for each
combination of disc parameters) to optimize the fit to the rms
velocity profile.  The best-fit model to the observed major-axis
stellar kinematics is obtained with a mass-to-light ratio \mlV$ =
3.42$ \mlsunV\ and with an exponential disc having a
surface-brightness profile
\begin{equation}
\mu_d(r) = 19.3 + 1.0857 \left(\frac{r}{22\farcs0}\right)\; 
{\rm mag\;arcsec}^{-2}, 
\end{equation}
(represented by the dashed line in the upper panel of
Fig.~\ref{fig_photometry}), a radial velocity dispersion profile
\begin{equation}
\sigma_R(r) = 155\;e^{-r/27\farcs4}\;
{\rm km\;s}^{-1},
\end{equation}
(where the galactocentric distance $r$ is expressed in arcsec) and an
inclination $i=72\degr$.  Fig.~\ref{fig_starmodel} shows the
comparison between the rotation curve, the dispersion velocity profile
and the rms velocity curve predicted by the best-fit model (solid
lines) with the observed stellar kinematics along the major axis of
NGC~4036. The agreement is good.

\begin{figure}
\centerline{\psfig{figure=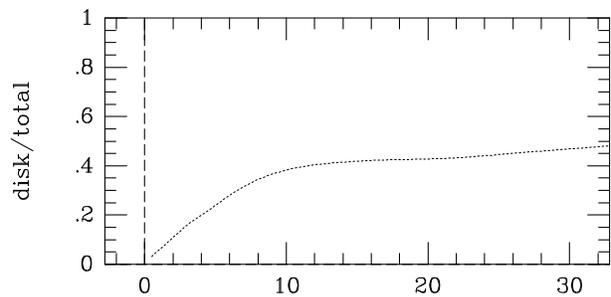,width=8cm}}
\caption{Radial profile of the fraction of the light contributed by the 
 stellar disc along the major axis of NGC~4036 as obtained by the best-fit 
 model. The distance from the center is expressed in arcsec.}
\label{fig_discratio}
\end{figure}

The derived $V-$band luminosities for bulge and disc are $L_b = 2.8
\cdot 10^{10}$ \lsunV\ and $L_d = 1.4 \cdot 10^{10}$ \lsunV. They
correspond to the masses $M_b = 9.8 \cdot 10^{10}$ \msun\ and $M_d =
4.8 \cdot 10^{10}$ \msun.  The total mass of the galaxy (bulge$+$disc)
is $M_T = 14.5 \cdot 10^{10}$ \msun .  The disc-to-bulge and
disc-to-total $V-$band luminosity ratios are $L_b/L_d=0.58$ and
$L_d/L_T=0.36$. The disc-to-total luminosity ratio as function of the
galactocentric distance is plotted in Fig.~\ref{fig_discratio}.

\subsubsection{Uncertainty ranges for the disc parameters}
\label{sec_uncertaintydisc}

\begin{figure}
\centerline{\psfig{figure=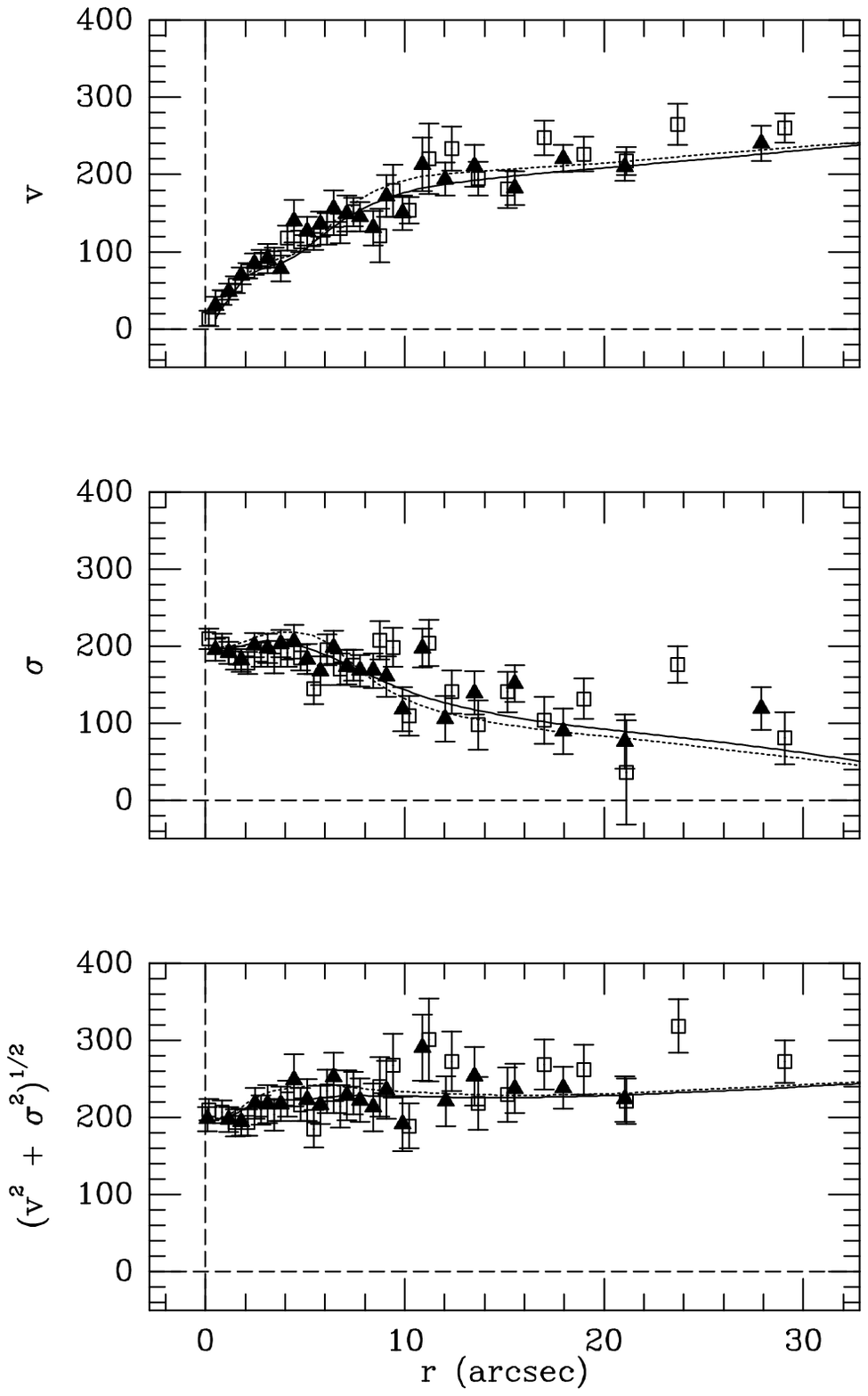,width=9.3cm,rwidth=6.7cm,rheight=13cm}}
\caption{As in Fig.~\ref{fig_starmodel} but for a disc with $\mu_0=19.1$ 
  \maq , $r_d = 20\farcs0$ and $i=72\degr$ (continuous line) and for a
  disc with $\mu_0=19.5$ \maq , $r_d = 24\farcs0$ and $i=72\degr$
  (dotted line).}
\label{fig_comparison1}
\end{figure}

\begin{figure}
\centerline{\psfig{figure=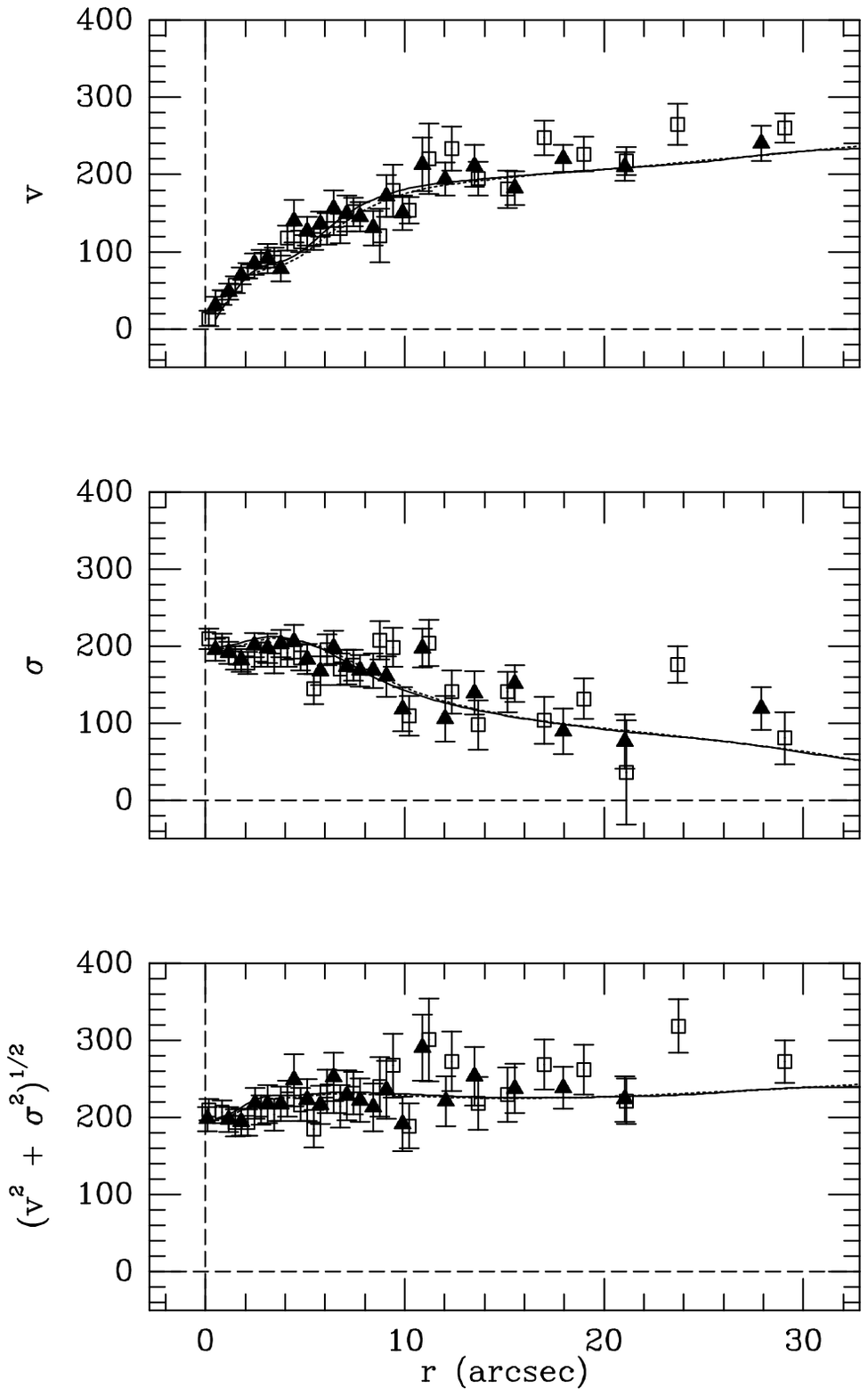,width=8cm}}
\caption{As in Fig.~\ref{fig_starmodel} but for a disc with $\mu_0=19.2$ 
  \maq , $r_d = 22\farcs0$ and $i=71\degr$ (continuous line) and for a
  disc with $\mu_0=19.4$ \maq , $r_d = 22\farcs0$ and $i=73\degr$
  (dotted line).}
\label{fig_comparison2}
\end{figure}

The uncertainty ranges for the disc parameters are $19.1 \la \mu_0
\la 19.5$ \maq, $20''\la r_d \la 24''$ and $71\degr \la i
\la 73\degr$.  In Fig.~\ref{fig_comparison1} the continuous and the
dotted line show the kinematical profiles predicted for two discs with
the same inclination ($i=72\degr$) and the same total luminosity of
the best-fit disc but with a smaller ($\mu_0=19.1$ \maq, $r_d =
20\farcs0$) or greater ($\mu_0=19.5$ \maq, $r_d = 24\farcs0$) scale
length respectively.  In Fig.~\ref{fig_comparison2} the continuous and
the dotted lines correspond to two discs with the same scale length
($r_d = 22\farcs0$) of the best-fit disc but with a lower
($\mu_0=19.2$ \maq, $i=71\degr$) or a higher ($\mu_0=19.4$ \maq, $i=73\degr$)
inclination.  Their total luminosity is respectively $\sim14\%$ lower
and $\sim16\%$ higher then that of the best-fit disc.  The fit of
these models to the data are also acceptable so we estimate a
$\sim15\%$ error in the determination of the disc luminosity and mass.
The good agreement with observations obtained with $k=1$,
$(M/L)_b=(M/L)_d$, and without a dark matter halo causes us to not
investigate models with $k\neq1$, $(M/L)_b\neq(M/L)_d$ or with
radially increasing mass-to-light ratios. Therefore we cannot exclude
them at all.

The $V-$band luminosity (after inclination correction) of the
exponential disc corresponding to the model best-fit to the observed
stellar kinematics is $\sim75\%$ than that of the disc obtained from
the parametric fit of the surface-brightness profiles. The differences
between the disc parameters derived from NGC~4036 photometry ($\mu_0 =
18.7$ \maq , $r_d = 22\farcs1$, $i=74\fdg9$) and from the stellar
kinematics ($\mu_0 = 19.3$ \maq , $r_d = 22\farcs0$, $i=72\degr$) are
expected.  In fitting the NGC~4036 surface-brightness profiles the
bulge is assumed (i) to be axisymmetric; (ii) to have a $R^{1/4}$
profile; (iii) and constant axial ratio (i.e. its isodensity
luminosity spheroids are similar concentric ellipsoids). We adopted
this kind of representation for the bulge component only to find rough
bounds on the exponential disc parameters to be used in the
kinematical fit. Often bulges have neither an $R^{1/4}$ law profile
(e.g. Burstein 1979; Simien \& Michard 1984) neither perfectly
elliptical isophotes (e.g. Scorza \& Bender 1990).  Therefore in
modeling the stellar kinematics the three-dimensional luminosity
density $j_b$ of the stellar bulge is assumed (i) to be oblate
axisymmetric; but (ii) not to be parametrized by any analytical
expression; (iii) nor to have isodensity luminosity spheroids with
constant axial ratio.  The flattening of the spheroids increases with
the galactocentric distance, as it appears from the three-dimensional
luminosity density profiles plotted in Fig.~\ref{fig_deprojection}
along different axes onto the meridional plane of NGC~4036.  This
flattening produces an increasing of the streaming motions for the
bulge component assumed to be an isotropic rotator.

\subsubsection{Modeling results with stellar kinematics by Fisher (1997)}
\label{sec_starmodelfisher}

\begin{figure}
\centerline{\psfig{figure=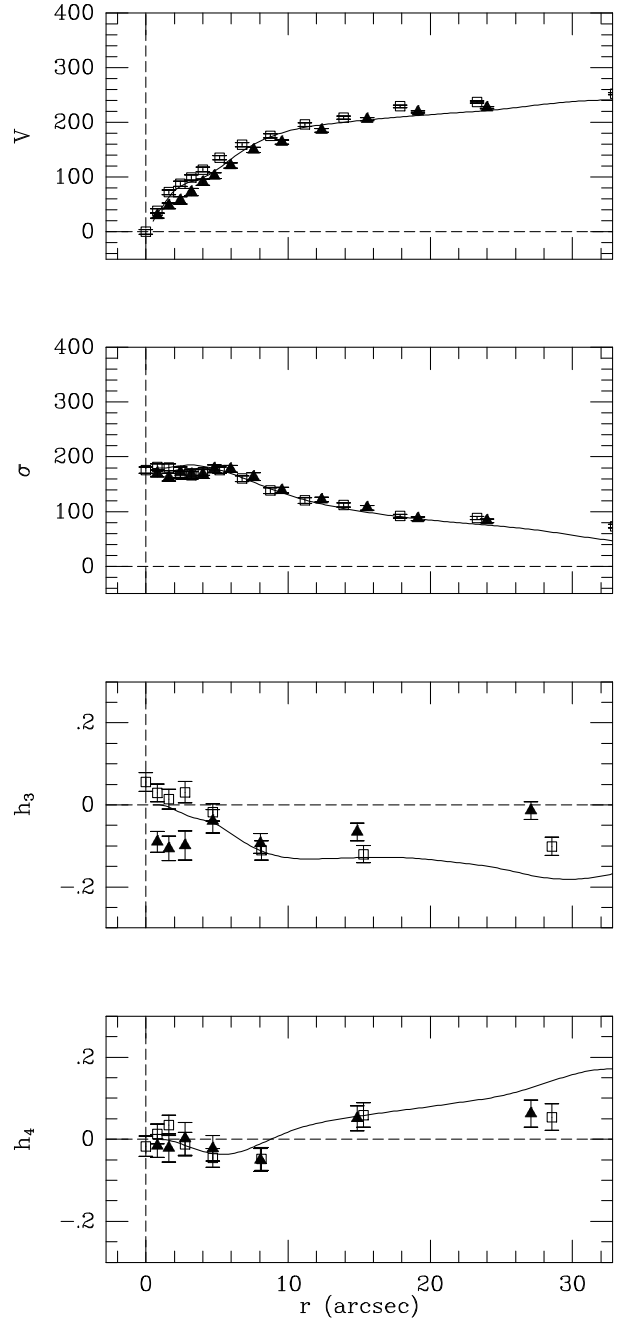,width=8truecm}}
\caption{Comparison between the predictions of the same model
  of Fig.~\ref{fig_starmodel} with the stellar major-axis kinematics
  for NGC 4036 by Fisher (1997). Data points are as in
  Fig.~\ref{fig_stars}. The solid curves represent the velocity $v$,
  velocity dispersion $\sigma$, $h_3$ and $h_4$ radial profiles of the
  best-fit model taking into account for Fisher's data reduction
  technique.}
\label{fig_starfisher}
\end{figure}

The bulge and disc parameters found for our best-fit model reproduce
also the major-axis stellar kinematics by F97.  Rather than choosing
an `ad hoc' wavenumber range to emulate F97 analysis technique the fit
to the predicted line profile was done in ordinary space and not in
the Fourier space as done to reproduce our Fourier Quotient
measurements.  The good agreement of the resulting kinematical
profiles with F97 measurements are shown in Fig.~\ref{fig_starfisher}.

\section{Modeling of the ionized gas kinematics}
\label{sec_gas}

\subsection{Modeling technique}
\label{sec_gasmodeling}

At small radii both the ionized gas velocity and velocity dispersion
are comparable to the stellar values, for $r\leq9''$ and $r\leq5''$
respectively. This prevented us to model the ionized gas kinematics by
assuming the gas had settled into a disc component as done by CvdM94
for NGC~2974.  Moreover a change in the slope of the \op\ intensity
radial profile (Fig.~\ref{fig_intensity}) is observed at $r\la8''$,
its gradient appears to be somewhat steeper towards the centre.  The
velocity dispersion and intensity profiles of the ionized gas suggest
that it is distributed into two components (see also the distinct
central structure in HST \ha$+$\nii\ image, Fig.~\ref{fig_halpha}): a
small inner spheroidal component and a disc.

We built a dynamical model for the ionized gas with a dynamically hot
spheroidal and in a dynamically colder disc component.

The total mass of the ionized gas $M_{\rm HII}$ is negligible and the
total potential is set only by the stellar component.  The mass of the
ionized gas can be derived from optical recombination line theory (see
Osterbrock 1989) by the \ha\ luminosity.  For a given electron
temperature $T_{\rm e}$ and density $N_{\rm e}$, the \hii\ mass is
given by
\begin{equation}
M_{\rm HII} = \left( L_{\rm H\alpha} m_{\rm H} /N_{\rm e} \right) /
\left( 4 \pi j_{\rm H\alpha} / N_{\rm e} N_{\rm p} \right)
\end{equation}
where \lha\ is the \ha\ luminosity, $m_{\rm H}$ is the mass of the
hydrogen atom, $j_{\rm H\alpha}$ is the \ha\ emissivity, and $N_{\rm
p}$ are the proton density (Tohline \& Osterbrock 1976).  The \ha\
luminosity of NGC~4036 scaled to the adopted distance is
\lha$\,=5.6\cdot10^{39}$ \lsun\ (Ho et al. 1997).  The
term $4 \pi j_{\rm H\alpha} / N_{\rm e} N_{\rm p}$ is insensitive to
changes of $N_{\rm e}$ over the range $10^2$ -- $10^6$ cm$^{-3}$.  It
decreases by a factor 3 for changes of $T_{\rm e}$ over the range
$5\cdot10^3\,\degr$K -- $2\cdot10^4\,\degr$K (Osterbrock 1989).  For
an assumed temperature $T_{\rm e}=10^4\,\degr$K, the electron density
is estimated to be $N_{\rm e}=2\cdot10^2$ cm$^{-3}$ from the \sii\
ratio found by Ho et al. (1997) implying $M_{\rm HII} = 7\cdot10^4$
\msun .

For the gaseous spheroid and disc we made two different sets of
assumptions based on two different physical scenarios for the gas
cloudlets.

\subsubsection{Long-lived gas cloudlets (model A)}
\label{sec_modelA}

In a first set of models we described the gaseous component by a set
of collisionless cloudlets in hydrostatic equilibrium.  The small
gaseous `spheroid' is characterized by a density distribution and
flattening different from those of stars.  Its major-axis luminosity
profile was assumed to follow an $R^{1/4}$ law. Adopting for it the
Ryden's (1992) analytical approximation, we obtained the
three-dimensional luminosity density of the gaseous spheroid.  The
flattening of the spheroids $q$ was kept as free parameter.  To derive
the kinematics of the gaseous spheroid we solved the Jeans Equations
under the same assumptions made in Sec.~\ref{sec_stars} for the
stellar spheroid. In particular the streaming velocity
$\overline{v_\phi}$ of gaseous bulge is derived from the second
azimuthal velocity moment $\overline{v_\phi^2}$ using Satoh's (1980)
relation.  For the ionized gaseous disc we solved the Jeans Equations
under similar assumptions made in Sec.~\ref{sec_stars} for the stellar
disc. Specifically, we assumed that the gaseous disc (i) has an
exponential luminosity profile; (ii) is infinitesimally thin; (iii)
has an exponentially decreasing $\sigma^2_R$; (iv) has $\sigma^2_z
=\sigma^2_R$; (v) has $\sigma^2_{\phi}$ satisfying the epicyclic
relation.

\subsubsection{Gas cloudlets `just' shed by the stars (model B)}
\label{sec_modelB}

In a second set of models we assumed that the emission observed in the
gaseous spheroid and disc arise from material that was recently shed
from stars. Different authors (Bertola et al. 1984; Fillmore et
al. 1986; Kormendy \& Westpfahl 1989; Mathews 1990) suggested that the
gas lost (e.g. in planetary nebulae) by stars was heated by shocks to
the virial temperature of the galaxy within $10^4$ years, a time
shorter than the typical dynamical time of the galaxy. Hence in this
picture the ionized gas and the stars have the same true kinematics,
while their observed kinematics are different due to the line-of-sight
integration of their different spatial distribution.  Differences
between the radial profiles of the gas emissivity and the stellar
luminosity may be explained if both the gas emission process and the
efficiency of the thermalization process show a variation with the
galactocentric distance. The three-dimensional luminosity density of
the spheroid is derived as in model A and the luminosity density
profile of the disc is assumed to be exponential.

In both cases, the kinematics of the gaseous spheroid and disc were
projected on the sky-plane to be compared to the observed ionized gas
kinematics. Assuming a Gaussian line profile for both the gaseous
components we derived the total line profile (which depends on the
relative flux of the two components).  As for the stellar components,
we convolved the line profiles obtained for the ionized gas with the
seeing PSF and we sampled them over the slit-width and pixel size to
mimic the observational setup. This procedure is particularly
important for the modeling of the observed kinematics near the
centre. As a last step (in mimicking the measuring technique of the
gaseous kinematics) we fitted a single Gaussian to the resulting line
profile, after taking into account for the instrumental line profile.

\subsection{Results for the gaseous component}
\label{sec_gasresults}

We decomposed the \op\ intensity profile as the sum of an $R^{1/4}$
gaseous spheroid and an exponential gaseous disc. A least-squares fit
of the observed data was done for $r>3''$ to deal with seeing effects
(Fig.~\ref{fig_intensity}).  The gas spheroid resulted to be the
dominating component up to $r\sim8''$ beyond the bright emission in
Fig.~\ref{fig_halpha}. We derived the effective radius of the gaseous
spheroid $r_{e,\,{\rm gas}} = 0\farcs5\pm0\farcs1$, the scale length
of the gaseous disc $r_{d,\,{\rm gas}} = 29\farcs8\pm0\farcs9$, and
the ratio between the effective intensity of the spheroid and the
central intensity of the disc $I_{e,\,{\rm gas}}/I_{0,\,{\rm gas}} =
718_{-153}^{+813}$.  The uncertainties on the resulting parameters
have been estimated by a separate decomposition on each side of the
galaxy.  With these parameters we applied the models for the gas
kinematics described in Sec.~\ref{sec_gasmodeling}.

Since in both models A and B the stellar density radial profile
differs from the gas emissivity radial profile, it is interesting to
check what is the relation between the three-dimensional stellar
density $\rho_{\rm star}(R)$ and the three-dimensional gas emissivity
$\nu_{\rm gas}(R)$. After deprojection we find they are related by the
following relation
\begin{equation}
\nu_{\rm gas}(R) \propto \rho_{\rm star}^{2.3}(R)
\label{eq_emissivity}
\end{equation}
in the range of galactocentric distances between $r\sim2''$ and
$r\sim10''$.  In a fully ionized gas the recombination rate is
proportional to the square of the gas density (e.g., Osterbrock 1989),
this is a power-law relation quite similar to Eq.~\ref{eq_emissivity}.

\begin{figure}
\centerline{\psfig{figure=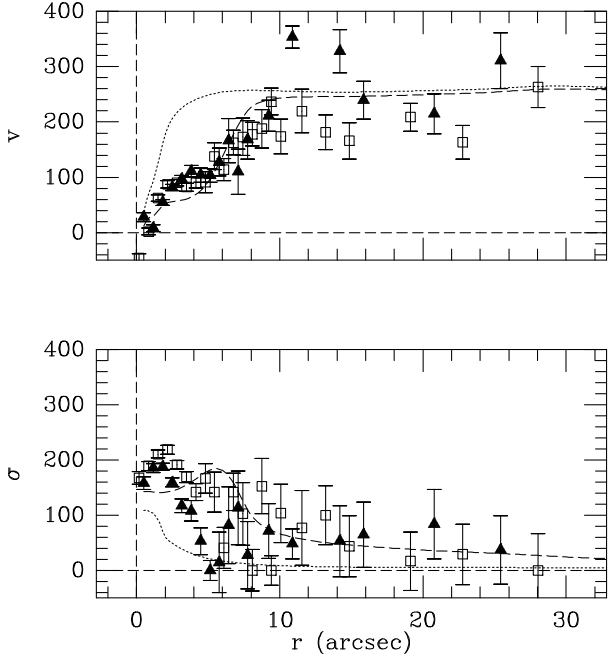,width=8cm}}
\caption{Comparison of the predictions of model A (dashed curve) 
  to the ionized gas kinematics observed along the major axis of
  NGC~4036.  The data points are as in Fig.~\ref{fig_gas}.  The dotted
  curves represent the seeing-convolved circular velocity curve and
  zero velocity-dispersion profile in the galaxy meridional plane.}
\label{fig_gasmodelA}
\end{figure}

\begin{figure}
\centerline{\psfig{figure=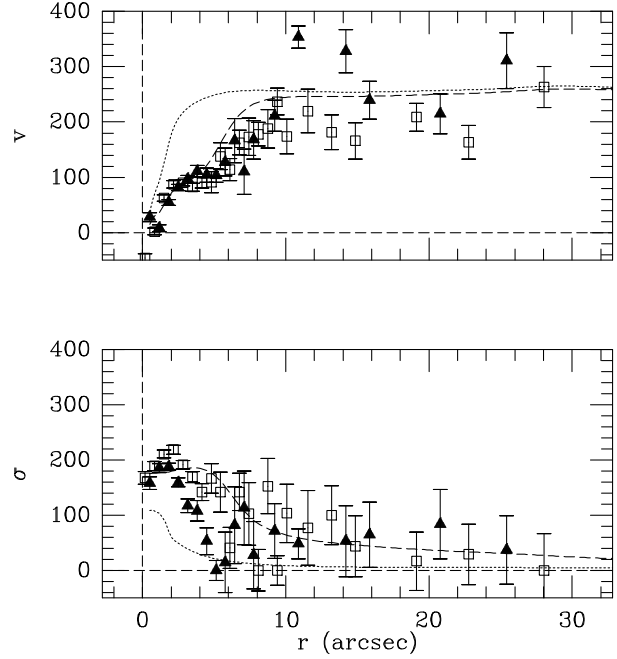,width=8cm}}
\caption{As in Fig.~\ref{fig_gasmodelA} but for model B.}
\label{fig_gasmodelB}
\end{figure}

For models A and B the best-fit to the observed gas kinematics are in
both cases obtained with a spheroid flattening $q=0.8$, and are
plotted respectively in Fig.~\ref{fig_gasmodelA} and
Fig.~\ref{fig_gasmodelB}.\\ 
The best result is obtained with model B. A simple estimate of the
errors on the model due to the uncertainties of bulge-disc
decomposition of the radial profile of
\op\ emission line can be inferred by comparing the model predictions
based on the separate decomposition on each side of the galaxy.  We
find a maximum difference of $5\%$ for $4''<r<10''$ between the gas
velocities and velocity dispersion predicted using the two different
bulge-disc decompositions of the \op\ intensity profile.  For model A
the assumption of Satoh's (1980) relation fails in reproducing the
observed gas kinematics for $r \la 6''$, where the emission lines
intensity profile is dominated by the gaseous spheroidal
component. However the $R^{1/4}$ extrapolation of the density profile
in the inner $3''$ overestimates the density gradient in this region
(as it appears also from the HST image, Fig.~\ref{fig_halpha}) which
could produce an exceeding asymmetric drift correction.

\subsubsection{Modeling results with gas kinematics by Fisher (1997)}
\label{sec_gasmodelfisher}

\begin{figure}
\centerline{\psfig{figure=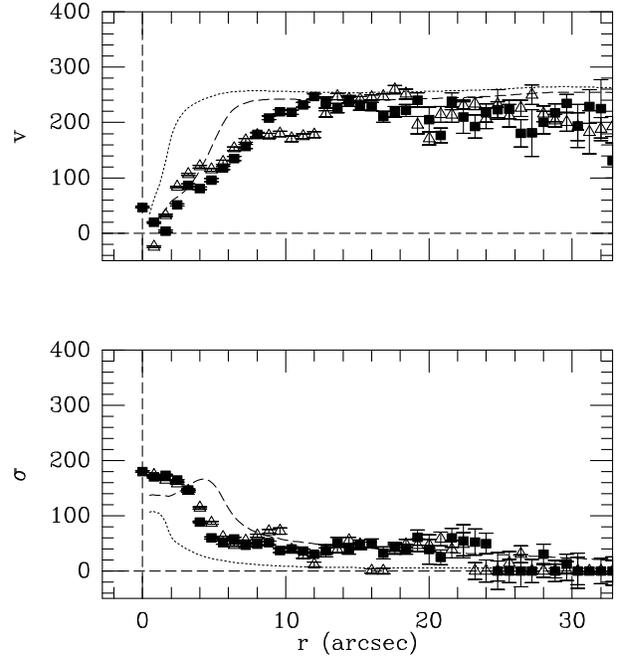,width=8cm}}
\caption{As in Fig.~\ref{fig_gasmodelA} but for the gas kinematics 
  measured by Fisher (1997). Open triangles and filled squares represent 
  data derived for the approaching W and receding E side
  respectively.} 
\label{fig_gasmodelAfisher}
\end{figure}

\begin{figure}
\centerline{\psfig{figure=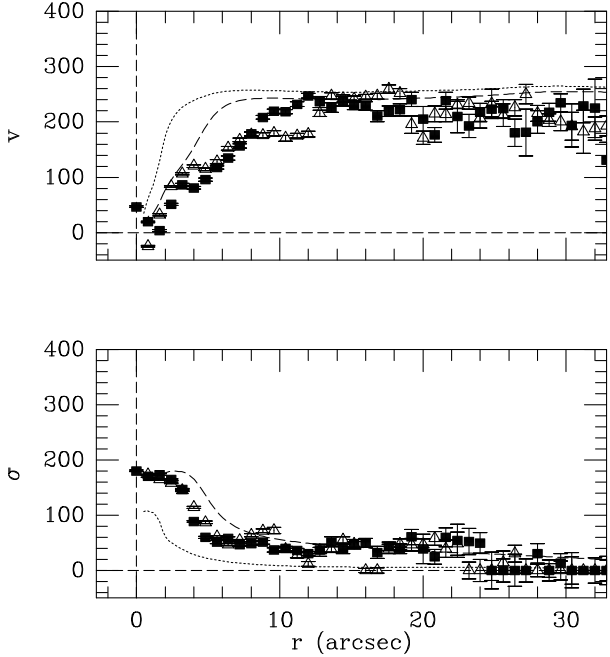,width=8cm}}
\caption{As in Fig.~\ref{fig_gasmodelAfisher} but for model B.}
\label{fig_gasmodelBfisher}
\end{figure}

We also applied our models A and B to the ionized gas kinematics and
to the \oiii\ intensity radial profile measured by F97 along the major
axis of NGC~4036. The best-fit to F97 data for models A and B are
obtained with a spheroid flattening $q=0.8$. They are shown in
Fig.~\ref{fig_gasmodelAfisher} and Fig.~\ref{fig_gasmodelBfisher}
respectively.

In this case the dynamical predictions of model B (even if they give
better results than those of model A) are not able to reproduce the
F97's kinematics in the radial range between $3''$ and $10''$. The
differences between the predicted and the measured kinematics rise up
to 80 \kms\ in velocities and to 40 \kms\ in velocity dispersions at
$r \sim 6''$. Nevertheless for $0''<r<10''$ this model agrees better
with the observations than if the gas is assumed to be on circular
orbits (see the dotted lines in Fig.~\ref{fig_gasmodelBfisher}).  If
this is the case the maximum differences with the measured kinematics
are large as 130 \kms\ in velocity and as 110 \kms\ in velocity
dispersion for $r \sim 4''$.

\subsection{Do drag forces affect the kinematics of the gaseous cloudlets?}
\label{sec_drag}

Considerable differences exist between the ionized gas kinematics
recently measured by F97 and the velocity and velocity dispersion
profiles predicted even by the best model in the bulge-dominated
region between $r\sim4''$ and $r\sim10''$.  This suggests that other
phenomena play a role in determining the dynamics of the spheroid gas
(see item (iii) in Sect.~\ref{sec_introduction}).  The discrepancy
between model and observations could be explained by accounting for
the drag interaction between the ionized gas and the hot component of
the interstellar medium.

In the scenario of evolution for stellar ejecta in elliptical galaxies
outlined by Mathews (1990), a portion of the gas shed by stars (e.g.,
as stellar winds or planetary nebulae) undergoes to an orbital
separation from its parent stars by the interaction with ambient gas,
after an expansion phase and the attainment of pressure equilibrium
with the environment, and before its disruption by various
instabilities.  Indeed to explain the luminosity of the optical
emission lines measured for nearby ellipticals, Mathews (1990)
estimated that the ionized gas ejected from the orbiting stars merges
with the hot interstellar medium in at least $t_{\rm life} \sim 10^6$
yr. If the gas clouds have at the beginning the same kinematics of
their parent stars, this lifetime is sufficiently long to let the
gaseous clouds (which start with the same kinematics of their parent
stars) to acquire an own kinematical behaviour due to the deceleration
produced by the drag force of the interstellar diffuse medium. The
lifetime of the ionized gas nebulae is shorter than $10^4-10^5$ years
if magnetics effects on gas kinematics are ignored (as in our model
B).

To have some qualitative insights in understanding the effects of a
drag force on the gas kinematics we studied the case of a gaseous
nebula moving in the spherical potential
\begin{equation}
\Phi(r) = \frac{4}{3}\, \pi\, G\, \rho\, r^2  
\label{eq_potential}
\end{equation}
generated by an homogeneous mass distribution of density $\rho$ and
which, starting onto a circular orbit, is decelerated by a drag force
\begin{equation}
{\bf F_{\it drag}} = - \frac{k_{\it drag}}{m}\, v^2\, \frac{{\bf v}}{v} 
\label{eq_drag}
\end{equation}
where $m$ and ${\bf v}$ are the mass and the velocity of the gaseous
cloud. Following Mathews (1990), the constant $k_{\it drag}$ is given by
\begin{equation}
k_{\it drag} \approx \frac{3}{4} \frac{n}{n_{\it eq}}\, \frac{m}{a_{\it
eq}} 
\end{equation}
where $n$ is the density of the interstellar medium, $n_{\it eq}$ and
$a_{\it eq}$ are respectively the density and the radius of the
gaseous nebula when the equilibrium is reached between the internal
pressure of the cloud and the external pressure of the interstellar
medium. The ratio $n/n_{\it eq} \sim 10^{-3}$ at any galactic radius
and therefore the ratio $k_{\it drag}/m$ depends on the nebula radius
$a_{\it eq}$ (Mathews 1990).

The equations of motion of the nebula expressed in plane polar
coordinates $(r, \psi)$ in which the centre of attraction is at $r=0$
and $\psi$ is the azimuthal angle in the orbital plane are
\begin{equation}
\ddot{r} - r\, \dot{\psi}^2 = 
- \frac{4}{3}\, \pi\, G\, \rho\, r + \frac{k_{\it drag}}{m}\, \dot{r}^2
\ (r>0)
\label{eq_moto1}
\end{equation}

\begin{equation}
r\, \ddot{\psi} + 2\, \dot{\psi}\, \dot{r} = 
- \frac{k_{\it drag}}{m}\, r^2\, \dot{\psi}^2 \ (r>0)
\label{eq_moto2}
\end{equation}

\begin{figure}
\centerline{\psfig{figure=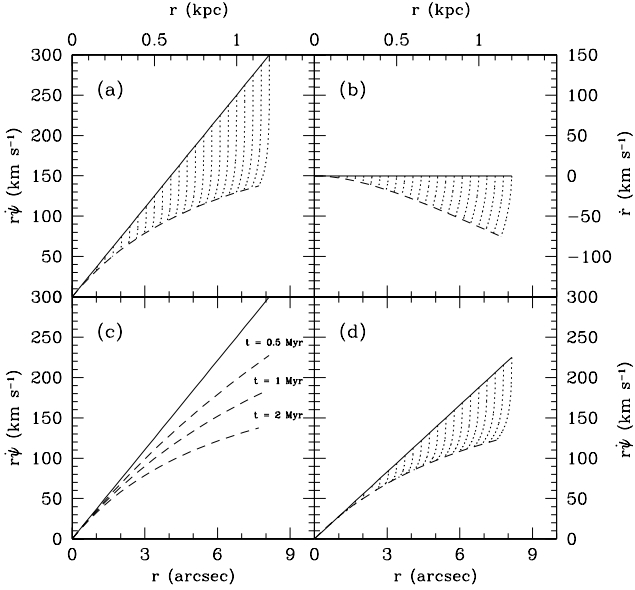,width=8.8cm}}
\caption{Variation of the tangential and radial components
  of the velocity of ionized gas nebulae which are initially supposed
  to move at circular velocity (continuous line) in the potential of
  the Eq.~\ref{eq_potential} under the effect of the drag force given
  by the Eq.~\ref{eq_drag}.  (a) and (b) The dashed lines represent
  the profiles of the tangential and radial components of the nebulae
  velocity as a function of the galactocentric distance after a time
  $t=2$ Myr.  (c) The radial profile of the velocity tangential
  component after after $t=0.5,\;1$ and $2$ Myr.  (d) The radial
  profile of the velocity tangential component after $t=2$ Myr for
  nebulae initially moving at a velocity 0.75 times the circular one.}
\label{fig_drag}
\end{figure}

We numerically solved the Eqs.~\ref{eq_moto1} and \ref{eq_moto2} with
the Runge-Kutta method (Press et al. 1986) to study the
time-dependence of the radial and tangential velocity components
$\dot{r}$ and $r\dot{\psi}$ of the nebula. We fixed the potential
assuming a circular velocity of 250 \kms\ at $r = 1$ kpc.  Following
Mathews (1990) we took an equilibrium radius for the gaseous nebula
$a_{\it eq}=0.37$ pc.  The results obtained for different times in
which the drag force decelerate the gaseous clouds are shown in
Fig.~\ref{fig_drag}.

It results that $\ddot{\psi}<0$ and $\ddot{r}>0$: the clouds spiralize
towards the galaxy centre as expected. Moreover the drag effects are
greater on faster starting clouds and therefore negligible for the
slowly moving clouds in the very inner region of NGC~4036.\\
If the nebulae are homogeneously distributed in the gaseous sphe\-roid
only the tangential component $r\dot{\psi}$ of their velocity
contribute to the observed velocity. No contribution derives from the
radial component $\dot{r}$ of their velocities. In fact for each
nebula moving towards the galaxy centre which is also approaching to
us we expect to find along the line-of-sight a receding nebula which
is falling to the centre from the same galactocentric distance with an
opposite line-of-sight component of its $\dot{r}$.

The radial components of the cloudlet velocities (typically of 30-40
\kms, see Fig.~\ref{fig_drag}) 
are crucial to explain the velocity dispersion profile and to
understand how the difference between the observed velocity
dispersions and the model B predictions arises. If the clouds are
decelerated by the drag force their orbits become more radially
extended and the velocity ellipsoids acquire a radial anisotropy.
This is a general effect and it is true not only in our case, in which
the clouds initially moved onto circular orbits.  So we expect (in the
region of the gaseous spheroid) the observed velocity dispersion
profile to decrease steeper than the one predicted by the isotropic
model B.  In fact inside $5'' \la r \la 10''$ the F97 ionized gas data
show that the gas velocity dispersion does not already exceed 50 \kms
, even if its rotation curve falls below the circular velocities
inferred from the stellar kinematics.  Given we do not know the the
lifetime and the density of the cloudlets we cannot make a definite
prediction.

The drag effects could explain the differences between the observed
and predicted velocities, if the decreasing of the tangential
velocities (showed in Fig.~\ref{fig_drag}) is considered as an upper
limit.\\
A proper luminosity-weighted integration of the tangential velocities
of nebulae along the line-of-sight has to be taken into account for
the gaseous spheroid. Being the clouds not all settled on a
particular plane, the plotted values will be reduced by two distinct
cosine-like terms depending on the two angles fixing the position of
any particular cloud.\\
Moreover (in the general case), the clouds are not all starting at the
local circular velocity and also at any given radius we find either
`younger' clouds (just shed from stars) and `older' clouds, which are
coming from farther regions of the gaseous spheroid and are going to
be thermalized.  Due to the nature of the drag force (acting much more
efficiently on fast-moving particles), it is easy to understand that
the fast `younger' clouds will soon leave their harbours, while the
slow `older' clouds will spend much more time crossing these regions.
For instance, in the case of nebulae shed at the local circular
velocity on a given plane (described by Eqs.~\ref{eq_moto1} and
\ref{eq_moto2} and shown Fig~\ref{fig_drag}) at $r=0.96$ kpc (where
the circular velocity is $\sim230$ \kms) we found that quite an half
of the gas cloudlets are coming from regions between 0.97 and 1.1 kpc
with tangential velocity between 150 \kms\ and 127 \kms .

This scenario is applicable only outside $3''$, whether if it is
applicable also inside the region of the central discrete structure
revealed by the HST image in Fig.~\ref{fig_halpha}, it is not clear.

\section{Discussion and conclusions}
\label{sec_conclusions}

The modeling of the stellar and gas kinematics in NGC~4036 shows that
the observed velocities of the ionized gas, moving in the
gravitational potential determined from the stellar kinematics, cannot
be explained without taking the gas velocity dispersion into
account. In the inner regions of NGC~4036 the gas is clearly not
moving at circular velocity. This finding is in agreement with earlier
results on other disc galaxies (Fillmore et al. 1986; Kent 1988;
Kormendy \& Westpfahl 1989) and ellipticals (CvdM94).

A much better match to the observed gas kinematics is found by
assuming the ionized gas as made of collisionless clouds in a
spheroidal and a disc component for which the Jeans Equations can be
solved in the gravitational potential of the stars (i.e., model A in
Sect.~\ref{sec_gasmodeling}).

Better agreement with the observed gas kinematics is achieved by
assuming that the ionized gas emission comes from material which has
recently been shed from the bulge stars (i.e., model B in
Sect.~\ref{sec_gasmodeling}). If this gas is heated to the virial
temperature of the galaxy (ceasing to produce emission lines) within a
time much shorter than the orbital time, it shares the same true
kinematics of its parent stars. If this is the case, we would observe
a different kinematics for ionized gas and stars due only to their
different spatial distribution.  The number of emission line photons
produced per unit mass of lost gas may depend on the environment and
therefore varying with the galactocentric distance. Therefore the
intensity radial profile of the emission lines of the ionized gas can
be different from that of the stellar luminosity. The
continuum-subtracted \ha$+$\nii\ image of the nucleus of NGC~4036
(Fig.~\ref{fig_halpha}) confirms that except for the complex emission
structure inside $\sim3''$ the smoothness of the distribution of the
emission as we expected for the gaseous spheroidal component.

In conclusion, the `slowly rising' gas rotation curve in the inner
region of NGC~4036 can be understood kinematically, at least in part.
The difference between the circular velocity curve (inferred from the
stellar kinematics) and the rotation curve measured for the ionized
gas is substantially due to the high velocity dispersion of the gas.\\
This kinematical modeling leaves open the questions about the physical
state (e.g. the lifetime of the emitting clouds) and the origin of the
dynamically hot gas.  We tested the hypothesis that the ionized gas is
located in short-living clouds shed by evolved stars (e.g. Mathews
1990) finding a reasonable agreement with our observational
data. These clouds may be ionized by the parent stars, by shocks, or
by the UV-flux from hot stars (Bertola et al. 1995a). The comparison
with the more recent and detailed data on gas by F97 opens wide the
possibility for further modeling improvement if the drag effects on
gaseous cloudlets (due to the diffuse interstellar medium) will be
taken into account.  These arguments indicate that the dynamically hot
gas in NGC~4036 has an internal origin. This does not exclude the
possibility for the gaseous disc to be of external origin as
discussed for S0's by Bertola, Buson \& Zeilinger (1992).

Spectra at higher spectral and spatial resolution are needed 
to understand the structure of the gas inside $3''$.
Two-dimensional spectra could further elucidate the nature of
the gas.

\section*{Acknowledgments}

We are indebted to Roeland van der Marel for providing for his
$f(E,L_{z})$ modeling software which became the basis of the programs
package used here.\\
WWZ acknowledges the support of the {\sl Jubil\"aumsfonds der
Oesterreichischen Nationalbank\/} (grant 6323).\\
This research made use of NASA/IPAC Extragalactic Database (NED) which
is operated by the Jet Propulsion Laboratory, California Institute of
Technology, under contract with the National Aeronautics and Space
Administration and of the Lyon-Meudon Extragalactic Database (LEDA)
supplied by the LEDA team at the CRAL-Observatoire de Lyon (France).

\appendix
\section{Data tables}

The stellar (Tab.~\ref{tab_stars}) and ionized gas
(Tab.~\ref{tab_gas}) heliocentric velocities and velocity
dispersions measured along the major axis of NGC~4036.

\begin{table}
\caption{Stellar kinematics along the major axis of NGC~4036}
\label{tab_stars}
\begin{tabular}{rcccc}
\multicolumn{1}{c}{$r$} & \multicolumn{1}{c}{$V$} &
\multicolumn{1}{c}{$\delta V$} & \multicolumn{1}{c}{$\sigma$} &
\multicolumn{1}{c}{$\delta\sigma$} \\
\multicolumn{1}{c}{[$''$]} & \multicolumn{1}{c}{[\kms]} &
\multicolumn{1}{c}{[\kms]} & \multicolumn{1}{c}{[\kms]} &
\multicolumn{1}{c}{[\kms]} \\
\multicolumn{1}{c}{(1)} & \multicolumn{1}{c}{(2)} & 
\multicolumn{1}{c}{(3)} & \multicolumn{1}{c}{(4)} & 
\multicolumn{1}{c}{(5)} \\
\\
$  -27.9$ &  1661 &   23 &   119 &   28 \\
$  -21.0$ &  1630 &   19 &    76 &   35 \\
$  -17.9$ &  1641 &   18 &    90 &   30 \\
$  -15.5$ &  1603 &   22 &   151 &   24 \\
$  -13.5$ &  1631 &   27 &   139 &   28 \\
$  -12.0$ &  1614 &   21 &   105 &   29 \\
$  -10.9$ &  1633 &   35 &   197 &   25 \\
$   -9.9$ &  1570 &   23 &   118 &   29 \\
$   -9.1$ &  1592 &   27 &   160 &   26 \\
$   -8.4$ &  1551 &   22 &   168 &   22 \\
$   -7.8$ &  1565 &   19 &   168 &   21 \\
$   -7.1$ &  1569 &   23 &   173 &   23 \\
$   -6.4$ &  1576 &   24 &   198 &   21 \\
$   -5.8$ &  1556 &   16 &   168 &   19 \\
$   -5.1$ &  1546 &   19 &   183 &   20 \\
$   -4.4$ &  1560 &   28 &   205 &   22 \\
$   -3.8$ &  1498 &   17 &   203 &   17 \\
$   -3.1$ &  1511 &   19 &   197 &   19 \\
$   -2.5$ &  1504 &   14 &   201 &   16 \\
$   -1.8$ &  1489 &   11 &   181 &   15 \\
$   -1.1$ &  1468 &   10 &   191 &   14 \\
$   -0.5$ &  1451 &   11 &   195 &   14 \\
$    0.2$ &  1407 &   10 &   209 &   13 \\
$    0.8$ &  1380 &    9 &   203 &   13 \\
$    1.5$ &  1363 &   11 &   185 &   15 \\
$    2.1$ &  1345 &   10 &   178 &   14 \\
$    2.8$ &  1330 &   12 &   189 &   16 \\
$    3.5$ &  1328 &   13 &   181 &   16 \\
$    4.1$ &  1302 &   16 &   191 &   18 \\
$    4.8$ &  1307 &   14 &   185 &   17 \\
$    5.4$ &  1304 &   14 &   144 &   20 \\
$    6.1$ &  1290 &   21 &   195 &   20 \\
$    6.8$ &  1290 &   19 &   170 &   21 \\
$    7.4$ &  1268 &   19 &   175 &   20 \\
$    8.7$ &  1299 &   34 &   207 &   25 \\
$    9.4$ &  1241 &   34 &   198 &   25 \\
$   10.2$ &  1266 &   17 &   110 &   26 \\
$   11.2$ &  1199 &   46 &   204 &   30 \\
$   12.4$ &  1186 &   28 &   141 &   28 \\
$   13.7$ &  1225 &   22 &    98 &   32 \\
$   15.2$ &  1239 &   24 &   140 &   26 \\
$   17.0$ &  1172 &   22 &   104 &   30 \\
$   19.0$ &  1193 &   23 &   131 &   26 \\
$   21.1$ &  1202 &   18 &    36 &   68 \\
$   23.7$ &  1155 &   26 &   176 &   24 \\
$   29.1$ &  1160 &   19 &    80 &   34 \\
\end{tabular}
\end{table}

\begin{table}
\caption{Ionized gas kinematics along the major axis of NGC~4036}
\label{tab_gas}
\begin{tabular}{rccccc}
\multicolumn{1}{c}{$r$} & \multicolumn{1}{c}{$V$} &
\multicolumn{1}{c}{$\delta V$} & \multicolumn{1}{c}{$\sigma$} &
\multicolumn{1}{c}{$\delta\sigma_+$} &
\multicolumn{1}{c}{$\delta\sigma_-$}\\
\multicolumn{1}{c}{[$''$]} & \multicolumn{1}{c}{[\kms]} &
\multicolumn{1}{c}{[\kms]} & \multicolumn{1}{c}{[\kms]} &
\multicolumn{1}{c}{[\kms]} & \multicolumn{1}{c}{[\kms]}\\
\multicolumn{1}{c}{(1)} & \multicolumn{1}{c}{(2)} & 
\multicolumn{1}{c}{(3)} & \multicolumn{1}{c}{(4)} & 
\multicolumn{1}{c}{(5)} & \multicolumn{1}{c}{(6)}\\
\\
$  -25.4$ &  1731 &   51 &    37 &   62 &  37 \\
$  -20.8$ &  1635 &   36 &    83 &   63 &  83 \\
$  -15.8$ &  1659 &   35 &    65 &   59 &  65 \\
$  -14.2$ &  1748 &   39 &    53 &   64 &  53 \\
$  -12.5$ &  1770 &   19 &     0 &    0 &   0 \\
$  -10.9$ &  1774 &   20 &    48 &   27 &  48 \\
$   -9.2$ &  1631 &   28 &    71 &   49 &  71 \\
$   -7.8$ &  1588 &   35 &    28 &   61 &  28 \\
$   -7.1$ &  1530 &   41 &   113 &   67 & 113 \\
$   -6.4$ &  1586 &   40 &    82 &   69 &  82 \\
$   -5.8$ &  1548 &   25 &    15 &   55 &  15 \\
$   -5.1$ &  1524 &   12 &     0 &   18 &   0 \\
$   -4.5$ &  1524 &   13 &    53 &   24 &  47 \\
$   -3.8$ &  1531 &   10 &   108 &   18 &  20 \\
$   -3.1$ &  1516 &    7 &   117 &   13 &  13 \\
$   -2.5$ &  1502 &    6 &   159 &    8 &   8 \\
$   -1.8$ &  1474 &    5 &   188 &    6 &   6 \\
$   -1.2$ &  1428 &    6 &   185 &    8 &   8 \\
$   -0.5$ &  1448 &    8 &   158 &   11 &  11 \\
$    0.2$ &  1466 &    8 &   167 &   12 &  12 \\
$    0.8$ &  1418 &    6 &   189 &    8 &   8 \\
$    1.5$ &  1358 &    6 &   211 &    7 &   7 \\
$    2.1$ &  1333 &    6 &   219 &    8 &   8 \\
$    2.8$ &  1330 &    6 &   191 &    8 &   8 \\
$    3.5$ &  1338 &    7 &   169 &   10 &  10 \\
$    4.1$ &  1330 &   10 &   142 &   15 &  16 \\
$    4.8$ &  1329 &   19 &   166 &   27 &  28 \\
$    5.4$ &  1282 &   24 &   141 &   36 &  39 \\
$    6.1$ &  1306 &   20 &    41 &   37 &  41 \\
$    6.8$ &  1257 &   23 &   142 &   34 &  36 \\
$    7.4$ &  1247 &   34 &   102 &   57 &  97 \\
$    8.1$ &  1242 &   22 &     0 &   38 &   0 \\
$    8.7$ &  1232 &   35 &   153 &   50 &  55 \\
$    9.4$ &  1183 &   24 &     0 &   27 &   0 \\
$   10.1$ &  1246 &   31 &   103 &   53 &  75 \\
$   11.6$ &  1200 &   39 &    77 &   67 &  77 \\
$   13.2$ &  1239 &   31 &   100 &   53 &  85 \\
$   14.9$ &  1254 &   33 &    43 &   55 &  43 \\
$   19.1$ &  1211 &   25 &    17 &   53 &  17 \\
$   22.8$ &  1256 &   31 &    29 &   55 &  29 \\
$   28.0$ &  1157 &   37 &     0 &   66 &   0 \\
\end{tabular}
\end{table}

\label{lastpage}


\clearpage

\begin{thebibliography}{99}

\bibitem{brn1} Bernacca P.L., Perinotto M., 1970, Contr. Oss. Astr.
Asiago,
	239, 1
\bibitem{ber1} Bertola F., Bettoni D., Rusconi L., Sedmak G., 
  	1984, AJ, 89, 356
\bibitem{ber5} Bertola F., Rubin V.C., Zeilinger W.W., 1989, ApJ, 345, L29
\bibitem{ber2} Bertola F., Buson L.M., Zeilinger W.W., 1992, ApJ, 
	401, L79
\bibitem{ber4} Bertola F., Bressan A., Burstein D., Buson L.M., 
	Chiosi C., di Serego Alighieri S., 1995a, ApJ, 438, 680
\bibitem{ber3} Bertola F., Cinzano P., Corsini E. M., Rix H.-W., 
	Zeilinger W.W., 1995b, ApJ, 448, L13
\bibitem{bet1} Bettoni D., Buson L.M., 1987, A\&AS, 67, 341
\bibitem{bin3} Binney J.J., Mamon G.A., 1982, MNRAS, 200, 361 
\bibitem{bin1} Binney J.J., Tremaine S., 1987, Galactic Dynamics. 
	Princeton University Press, Princeton
\bibitem{bin2} Binney J.J., Davies R.L., Illingworth, G.D., 1990, 
	ApJ, 361, 78
\bibitem{bur1} Burstein, D., 1979, ApJ, 234, 829
\bibitem{cin1} Cinzano P., 1995, PhD thesis, Universit\`a di Padova
\bibitem{cin2} Cinzano P., van der Marel R.P., 1994, MNRAS, 270, 325
          (CvdM94)
\bibitem{cor1} Corsini E.M., Pizzella A., Sarzi M., Cinzano P., Vega
          Beltr\'an J.C., Funes J.G., Bertola F., Persic M., Salucci P., 
          1998, A\&A, in press [astro-ph/9809366]
\bibitem{cud1} Cuddeford P., Binney J.J., 1994, MNRAS, 266, 273
\bibitem{dev1} de Vaucouleurs G., de Vaucouleurs A., Corwin H.G.Jr., 
	Buta R.J., Paturel H.G., Fouqu\'e P., 1991, Third Reference 
	Catalogue of Bright Galaxies. Springer-Verlag, New York (RC3)
\bibitem{eva2} Evans D.S., 1967, in Batten A.H., Heard J.F., eds,
	Proc. IAU Symp. 30, Determination of Radial Velocities and their 
	Applications. Academic Press, London, p.~57
\bibitem{eva1} Evans N.W., Collett J.L., 1993, MNRAS, 264, 353
\bibitem{fas1} Fasano G., 1990, Internal Report,  
	Astronomical Observatory, Padova
\bibitem{fill1} Fillmore J.A., Boroson T.A., Dressler A., 1986, ApJ, 
	302, 208
\bibitem{fis1} Fisher D., 1997, AJ, 113, 950 (F97)
\bibitem{gar1} Garcia A.M., 1993, A\&AS, 100, 47  
\bibitem{ger1} Gerhard O.E., 1993, MNRAS, 265, 213
\bibitem{ho1}  Ho L.C., Filippenko A.V., Sargent W.L.W, 1997, ApJS, 112,
         315 
\bibitem{jaf1} Jaffe W., 1983, MNRAS, 202, 995
\bibitem{ken1} Kent S.M., 1984, ApJS, 56, 105
\bibitem{ken2} Kent S.M., 1985, ApJS, 59, 115
\bibitem{ken3} Kent S.M., 1988, ApJ, 96, 514
\bibitem{kor1} Kormendy J., Westpfahl D.J, 1989, ApJ, 338, 752
\bibitem{kui2} Kuijken K., Merrifield, M.R., 1993, MNRAS, 264, 712
\bibitem{kui1} Kuijken K., Tremaine S., 1992, in Sundelius B., ed.,
	Dynamics of Disk Galaxies. University of Goteborg Press, 
	Goteborg, p.~71
\bibitem{luc1} Lucy L.B., 1974, AJ, 79, 745
\bibitem{mat1} Mathews W.G., 1990, ApJ, 354, 468
\bibitem{mic1} Michard R., 1993, in  Danziger I.J., Zeilinger W.W., 
	Kj\"ar K., eds, Structure, Dynamics and Chemical Evolution of 
	Elliptical Galaxies. ESO, Garching, p. 553
\bibitem{ost1} Osterbrock D.E., 1989, Astrophysics of Gaseous Nebulae and
       	Active Galactic Nuclei. University Science Books, Mill Valley
\bibitem{pel1} Peletier R.F., Davies R.L., Illingworth G.D., Davies L.E.,
    	Cawson M., 1990, AJ, 100, 1091
\bibitem{pre1} Press W.H., Flannery B.P., Teukolsky S.A., Vetterling W.T.,
	1986, Numerical Recipes. Cambridge University Press, Cambridge
\bibitem{ric1} Richardson M.B., 1972, J. Opt. Soc. Am., 62, 55
\bibitem{rix1} Rix H.-W., White S.D.M., 1992, MNRAS, 254, 389
\bibitem{rix2} Rix H.-W., Kennicutt R.C.Jr., Braun R., Walterbos R.A.M., 
	1995, ApJ, 438, 155
\bibitem{rix3} Rix H.-W., de Zeeuw T., Cretton N., van der Marel R.P., 
         Carollo M., 1997, ApJ, 488, 702
\bibitem{rob1} Roberts M.S., Hogg D.E., Bregman J.N., Forman W.R., 
	Jones C.R., 1991, ApJS, 75, 751
\bibitem{ryd1} Ryden B., 1992, ApJ, 386, 42
\bibitem{san1} Sandage A., Bedke J., 1994, The Carnegie Atlas of Galaxies.
    	Carnegie Institution, Flintridge Foundation, Washington (CAG)
\bibitem{san2} Sandage A., Tammann G.A., 1981, A Revised Shapley--Ames
    	Catalog of Bright Galaxies. Carnegie Institution, Washington (RSA)
\bibitem{sar1} Sargent W.L.W., Schechter P.L., Boksenberg A., 
	Shortridge K., 1977, ApJ, 212, 326
\bibitem{sat1} Satoh C., 1980, PASJ, 32, 41
\bibitem{sco1} Scorza C., Bender R., 1990, A\&A, 235, 49
\bibitem{sim1} Simien F., Michard R., 1984, in Nieto J.-L., ed., 
	New Aspects of Galaxy Photometry. Springer, Berlin, p.~345
\bibitem{toh1} Tohline J.E., Osterbrock D.E., 1976, ApJ, 210, L117
\bibitem{vdm1} van der Marel R.P., 1991, MNRAS, 253, 710
\bibitem{vdm2} van der Marel R.P., Cinzano P., 1992, in
    	Busarello G., Capaccioli M., Longo G., eds, Morphological and 
	Physical Classification of Galaxies. Kluwer, Dordrecht, p.~437
\bibitem{vdm3} van der Marel R.P., Franx M., 1993, ApJ, 407, 525
\bibitem{vdm6} van der Marel R.P., Binney J.J., Davies R.L., 1990, MNRAS,
	245, 582
\bibitem{vdm4} van der Marel R.P., Rix H.-W., Carter D., Franx M., 
	White S.D.M., de Zeeuw P.T., 1994, MNRAS, 268, 521
\bibitem{whi1} White R.L., 1994, in Crabtree D.R., Hanisch R.J., Barnes
          J., eds, ASP Conf. Ser. 61, Astronomical Data Analysis Software 
	and Systems III. ASP, San Francisco, p.~292
\end{thebibliography}
\end{document}